\documentclass[journal]{IEEEtran}
\usepackage{graphicx}
\usepackage[caption=false,font=normalsize,labelfont=sf,textfont=sf]{subfig}
\usepackage{cite}
\usepackage{picinpar}
\usepackage{amsmath}
\usepackage{url}
\usepackage[utf8]{inputenc}
\usepackage{colortbl}
\usepackage{soul}
\usepackage{multirow}
\usepackage{pifont}
\usepackage{psfrag}
\usepackage{alltt}
\usepackage[hidelinks]{hyperref}
\usepackage{enumerate}
\usepackage{siunitx}
\usepackage{breakurl}
\usepackage{epstopdf}
\usepackage{pbox}
\usepackage[usenames,dvipsnames,table]{xcolor}
\usepackage{enumitem}

\usepackage{amssymb}
\usepackage{verbatim}

\begin{document}
%
% paper title
% Titles are generally capitalized except for words such as a, an, and, as,
% at, but, by, for, in, nor, of, on, or, the, to and up, which are usually
% not capitalized unless they are the first or last word of the title.
% Linebreaks \\ can be used within to get better formatting as desired.
% Do not put math or special symbols in the title.
%\title{Hairpin Windings---\textcolor{red}{Reintroduction?} of a Lost Art}

%\title{Hairpin Windings: Twists and Bends of a Technological Breakthrough that Arrived A Century Too Soon}
%\marginpar{Raiders of the Lost Art}
% Revival of an Old Technology
\title{Hairpin Motors for Electromobility: Twists and Bends of a Technological Breakthrough that Initially Arrived A Century Too Soon}

% author names and IEEE memberships
% note positions of commas and nonbreaking spaces ( ~ ) LaTeX will not break
% a structure at a ~ so this keeps an author's name from being broken across
% two lines.
% use \thanks{} to gain access to the first footnote area
% a separate \thanks must be used for each paragraph as LaTeX2e's \thanks
% was not built to handle multiple paragraphs
%

\author{Stefan~M.~Goetz,
        Ricardo~Lizana~F.,
        and~Sebastian~Rivera
        % <-this % stops a space
\thanks{This work was supported in part by Trinity College's Isaac Newton Trust Fund, the Duke Energy Initiative, KSB Foundation, and the NSF project No.\ 1608929. Also, financial support by the following projects from the Agencia Nacional de Investigación y Desarrollo (ANID): FONDECYT Grant 1220346, REDES-PCI 190108, AC3E (ANID/Basal/FB0008), SERC Chile (ANID/FONDAP/1522A0006) and ANID/EQM/180215. Finally, the support from the Fondo de Actividades Academicas (FAA) from Universidad Catolica de la Santisima Concepcion is greatly acknowledged.

S.~M.~Goetz is with the University of Cambridge, CB3 0FA, UK, and Duke University, Durham, NC 27710, USA (e-mail: smg84@cam.ac.uk).

R.~Lizana and F.~Figueroa are with the Department of Electrical Engineering, Universidad Cat\'olica de la Sant\'isima Concepci\'on, Centro de Energ\'ia, Concepci\'on, Chile (e-mail: ricardolizana@ucsc.cl).

S.~Rivera is with the Department of Electrical Sustainable Energy, DCE\&S group, Delft University of Technology, 2628 CD Delft, South Holland, The Netherlands and also with the Department of Electrical Engineering, Universidad Cat\'olica de la Sant\'isima Concepci\'on, Centro de Energ\'ia, Concepci\'on 4090541, Chile (e-mail: s.rivera.i@ieee.org).

}
}

% The paper headers
\markboth{Scanning Our Past}%
{Shell \MakeLowercase{\textit{et al.}}: Bare Demo of IEEEtran.cls for IEEE Journals}
% The only time the second header will appear is for the odd numbered pages
% after the title page when using the twoside option.
% 
% *** Note that you probably will NOT want to include the author's ***
% *** name in the headers of peer review papers.                   ***
% You can use \ifCLASSOPTIONpeerreview for conditional compilation here if
% you desire.

% If you want to put a publisher's ID mark on the page you can do it like
% this:
%\IEEEpubid{0000--0000/00\$00.00~\copyright~2015 IEEE}
% Remember, if you use this you must call \IEEEpubidadjcol in the second
% column for its text to clear the IEEEpubid mark.

% use for special paper notices
%\IEEEspecialpapernotice{(Invited Paper)}

% make the title area
\maketitle

% As a general rule, do not put math, special symbols or citations
% in the abstract or keywords.
\begin{abstract}
There is currently a major trend to hairpin-winding motors for small and medium drives with increased power, specifically more torque density in the automotive industry. Practically all large players in the field either already use this winding technology or have announced doing so soon. However, hairpins, bar windings, and other segmented winding techniques are not purely a material and production issue. Instead their application to small drives influences all aspects of the design of machines, which are currently explored and studied by the industry. These range from not obvious gaps in the theory, parameter studies for maxima of efficiency, possible as well as advantageous winding schemes, thermal design, and ways to control ac losses to specific materials and process difficulties.

Despite the apparent novelty of the trend, however, designers could revisit a widely forgotten knowledge base of more than 100 years for many of those questions. This old knowledge base and the understanding that many recently presented concepts have been developed earlier may speed up the technological development and appear to be a key to further innovation. Instead, many problems need to be solved again and technologies re-invented. Furthermore, as this technology has recently become merely industry-driven, a substantial portion of the information and technological developments are not available to the public---a state that to our eyes may harm the innovation capacity of the drives community. \end{abstract}

% Note that keywords are not normally used for peerreview papers.
%\begin{IEEEkeywords}IEEE, IEEEtran, journal, \LaTeX, paper, template.\end{IEEEkeywords}

% For peer review papers, you can put extra information on the cover
% page as needed:
% \ifCLASSOPTIONpeerreview
% \begin{center} \bfseries EDICS Category: 3-BBND \end{center}
% \fi
%
% For peerreview papers, this IEEEtran command inserts a page break and
% creates the second title. It will be ignored for other modes.
\IEEEpeerreviewmaketitle

%3.1.5. Producing preformed Coils

\section{Introduction}
% The very first letter is a 2 line initial drop letter followed
% by the rest of the first word in caps.
% 
% form to use if the first word consists of a single letter:
% \IEEEPARstart{A}{demo} file is ....
% 
% form to use if you need the single drop letter followed by
% normal text (unknown if ever used by the IEEE):
% \IEEEPARstart{A}{}demo file is ....
% 
% Some journals put the first two words in caps:
% \IEEEPARstart{T}{his demo} file is ....
% 
% Here we have the typical use of a "T" for an initial drop letter
% and "HIS" in caps to complete the first word.

\begin{figure}[t!]
 \centering{
 \subfloat[]{\includegraphics[height=0.2\textwidth]{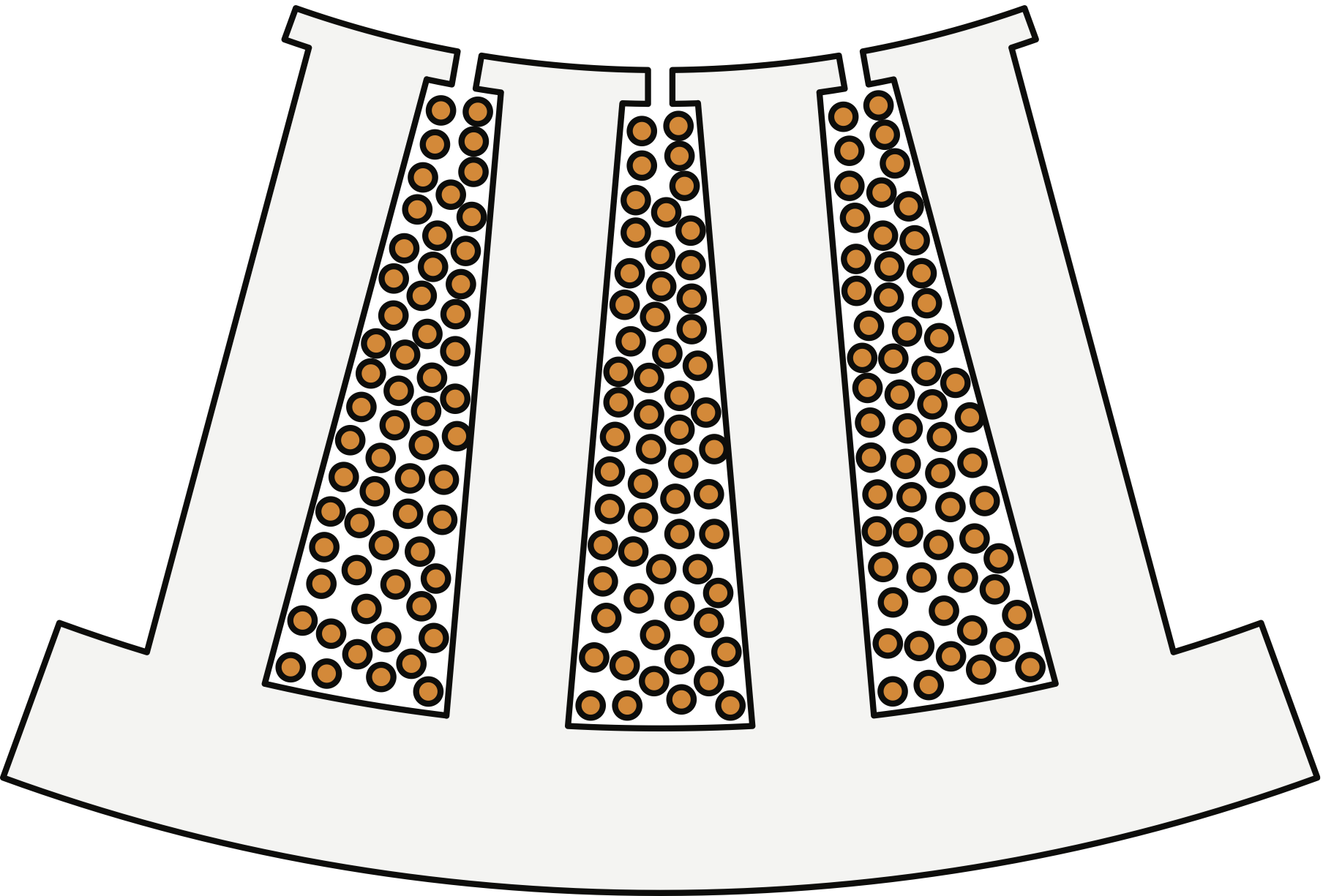}\label{round_w}}\hfill
 \subfloat[]{\includegraphics[height=0.2\textwidth]{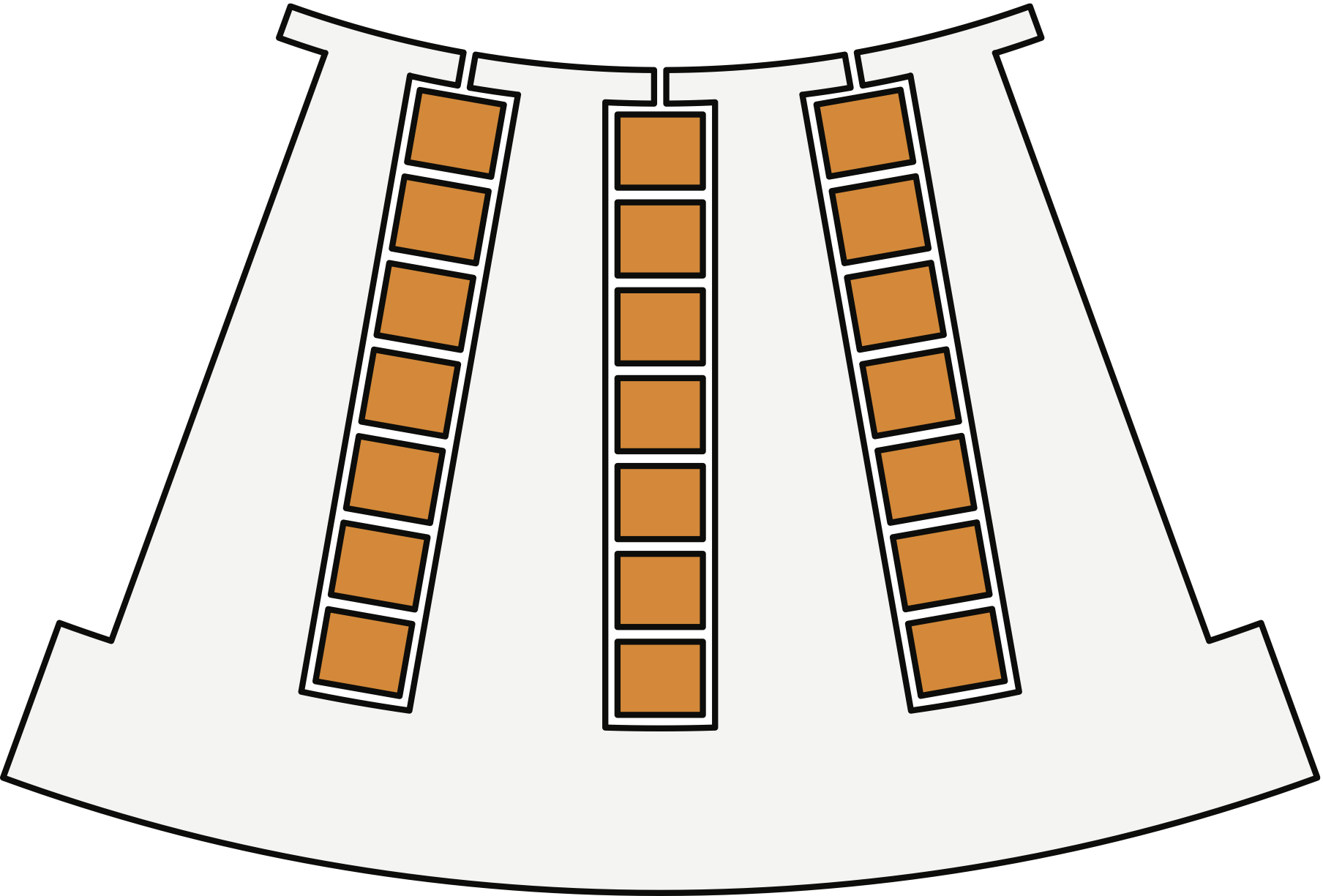}\label{bar_w}}
 \caption{Comparison of (a) wire-wound and (b) bar-wound machines illustrated with three stator slots. In the wire-wound case, the cross-section of the teeth is determined by the available space near the air gap, which also determines the flux saturation, and typically stays the same in radial direction to assign the to the slot and the winding. Bar-wound machines at present typically use rectangular slots and equal bar size in radial direction. The remaining space is allotted to the teeth. Saturation would set in near the air gap accordingly.}\label{windings}
 }
 \end{figure}

\IEEEPARstart{M}{agnetic windings} in general and subsequently of small drives are typically associated with thin round copper wires. This group of small drives includes electrical machines for automotive applications, ranging from ancillary units to traction machines for both hybrid-electric and battery-electric vehicles \cite{Hughes:2019,Hendershot:2010}. Wire-wound machines can refer to well-established techniques for widely automatic manufacturing---with the exception of traction machines with distributed windings, which still contain manual steps in most assembly lines, particularly after the insertion process \cite{HagedornBook}.
%Also, despite the well-established processes for industrial electric motors, these are not suitable for large-scale productions in automotive industry.
The final position of each wire in the slots of such wire-wound machines is typically not controlled, which can have advantages with respect to high-frequency effects and led to the alternative name of random wire wound, as displayed in Fig.\ \ref{windings}\subref{round_w} \cite{doi:https://doi.org/10.1002/9781118701591.ch2}.

Since the first larger series use in a drive train, windings made of thick copper bars or strips with rectangular profiles have rapidly gained momentum in the automotive world \cite{RahmanSAEI, 2015-01-1208, 8658353,9541356}. Such bars with well-controlled position in the slot enable larger cross sections for higher currents, in addition to a better fit to the likewise rectangular slots with clearly defined geometries, as presented in Fig. \ref{windings}\subref{bar_w}. The matched surfaces also increase the thermal contact area between conductors and slot walls. For easier insertion and due to the higher stiffness, the entire winding is split into smaller segments that are largely shaped beforehand. Most dominant are U-shaped segments, which in accordance with their appearance are widely called hairpins.

\begin{figure}[t!]
\centering{
 \subfloat[]{\includegraphics[scale=.054]{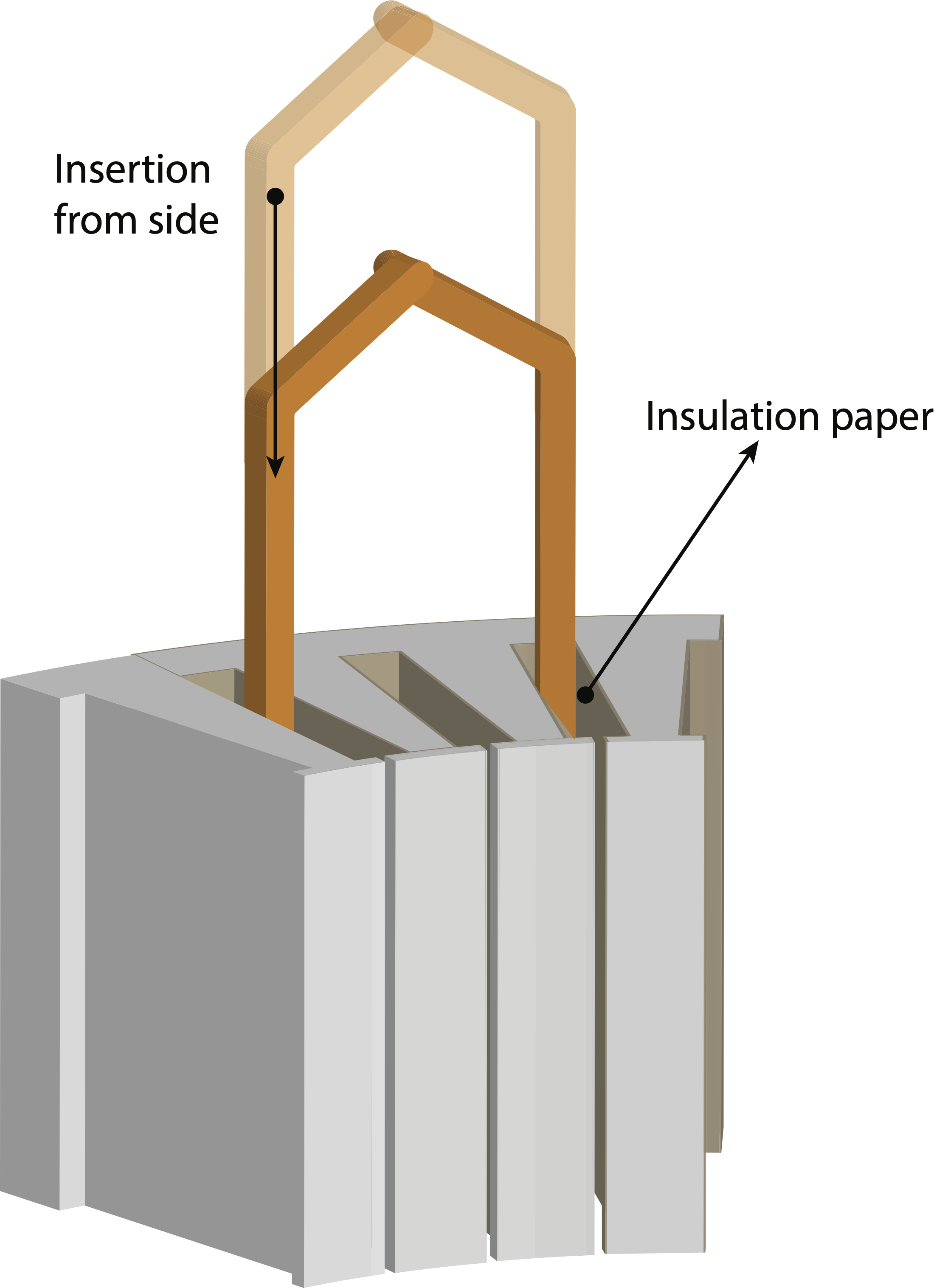}\label{side}} \hskip 0.2cm %\hfill
 \subfloat[]{\includegraphics[scale=.054]{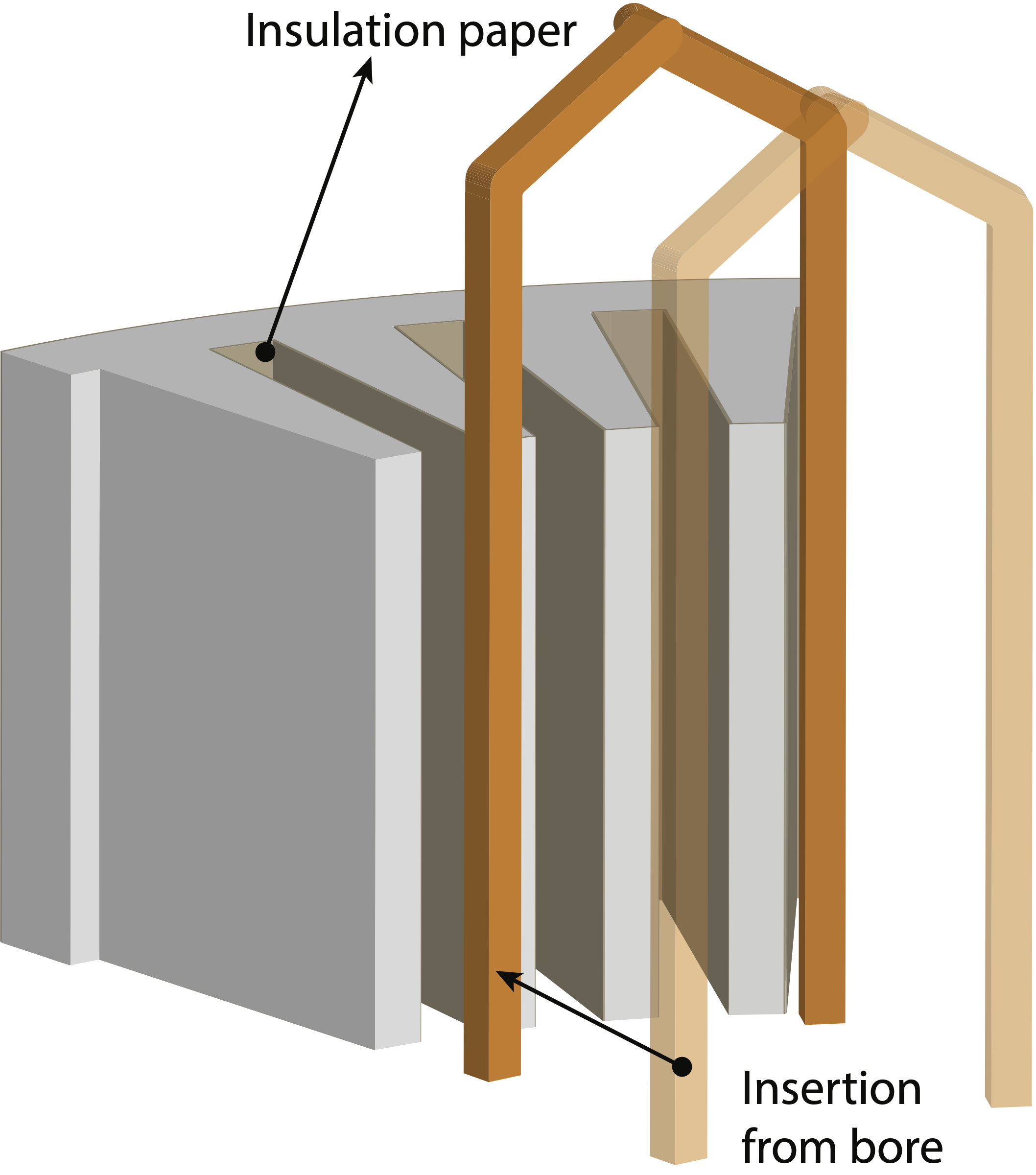}\label{bore}}
\caption{Hairpin insertion options: (a) from the end-turn side, allowing (partially) closed slots but requiring short segments (I or U pins) and subsequent welding or (b) from the bore requiring open slots but enabling longer segments.}\label{hairpin}
}
\end{figure}

This intermediate U shape typically already has the in-slot part of the arms aligned with the specific stator's slot walls (hence, the cross sections of the two legs are slightly twisted against each other to account for the different orientation of the two slots into which they will be inserted) and matches not only the coil pitch but also the radius of the bar position to the shaft axis. Furthermore, the hairpins already form the overhang, \emph{i.e.}, the end-turns on one side with their cusps. Thus, these conductor segments are close to the final shape in the machine and, on the other hand, as displayed in Fig.\ \ref{hairpin} can still be inserted easily into the slots either from the front side or alternatively from the bore in case the slots are entirely open and do not have overhanging teeth \cite{7008430,8766938}. After insertion into the slots, the individual segments have to be electrically connected subsequently to form the final continuous winding. %Trend: thicker, but rectangular copper segments provide larger cross section for higher currents, fit perfectly into slot, but too stiff for winding => segmented and largely shaped before insertion; have to be electrically connected later

The sudden success of bar windings is credited to a number of advantages. Due to the growing quantities of electric vehicles, full automation particularly of hairpin designs offers a clear substantial cost as well as quality advantage in mass production \cite{8658331,RahmanSAEI,2015-01-1208,7310394,7409310,Weigelt2018,Acatech2010, Hawkins2014, Rahman2016}. Furthermore, up to a factor of two increase in slot fill in addition to 50\% reduction of end-turn length enable higher power densities \cite{8766938}. The better defined and also larger interfaces between the rectangular conductors and the slot walls improve heat dissipation to further increase the achievable power density \cite{7008430,7310394,8766938,7296667}. Beyond the heat conduction to the slot walls, the large surface area and the clearance between the conductors at the end turns additionally allow highly effective direct cooling through liquid perfusion of this braid-like end-turn structure (see the section Latest Developments). The larger bars also substantially increase the possible insulation thickness as well as resistance, lifetime, and reproducibility \cite{Stone:2007,7604085,2016-01-1228,8658353}. Better reproducibility is a general feature of bar-wound machines and can extend to diagnosis, rework, and repair, although the latter may be more relevant in small-series vehicles and/or during development.

At present, automotive traction motors aggressively increase the torque density and the voltage, resulting in power densities and temperature levels that conventional industrial drives are not able to achieve for reliability and cost reasons \cite{7873375, 9316773, 7917257}. Bar-winding technology, \emph{e.g.}, with hairpins, appears to be a good match and enabler for these goals.

Interestingly, most of these features were already demonstrated and promoted more than a century ago. However, at that time, bar windings in general and hairpin designs in particular were strongly limited by ac losses and initially also rudimentary manufacturing automation \cite{Krause1910}. Since then, bar-wound machines have mostly covered a niche in low-voltage high-current applications. Whereas the trend of the automotive industry towards electric drive trains generates needs almost perfectly met by bar windings, there are also new challenges.
%, leading to larger voltages in a smaller packages.

%The strong technological and quantitative demand driven by the electrification of automotive power trains along with the trend towards \textit{more electric} solutions elsewhere have catalyzed the development of new technologies and production processes for electric (traction) motors \cite{7323198,Weigelt2018,8658353,7296667}. Significant advancements have lead to highly efficient manufacturing processes which not only match the automotive production standards: fully automated, resource-efficient with short lead times, but also enable additional benefits that are decisive in this challenging industry \textbf{\cite{Acatech2010}: improved power density and efficiency, larger slot fill factor (up to 80\%), simpler heat dissipation, smaller package space, reduction of Joule losses, smaller winding overhang, or extended insulation lifetime to name a few \cite{7008430,7604085,8658353,8766938}.}

%\section{Historical Development}
%\section{How it all started}
\section{Historical Development}
Already the earliest electrical machines as well as transformers have primarily used round electrical wires in their structure \cite{US422746A,US593138,doi:10.1080/14786446808640056, 8068865}. However, designers quickly recognized the potential of higher packing and slot fill through the use of conductor bars. Carl William Siemens, for instance, discussed this very topic in 1880 as a means to reduce the electrical resistance for dynamo machines in one of his brother Werner's machines \cite{siemens1883science,Siemens:1919, US264780A}. %Siemens:1919 p.451
The long known low resistance of bar conductors of copper or bronze is also appreciated in the contemporary patent literature \cite{Bradley:1988}.

\begin{figure*}[tb]
    \centering{
    \subfloat[]{\includegraphics[height=0.35\textwidth]{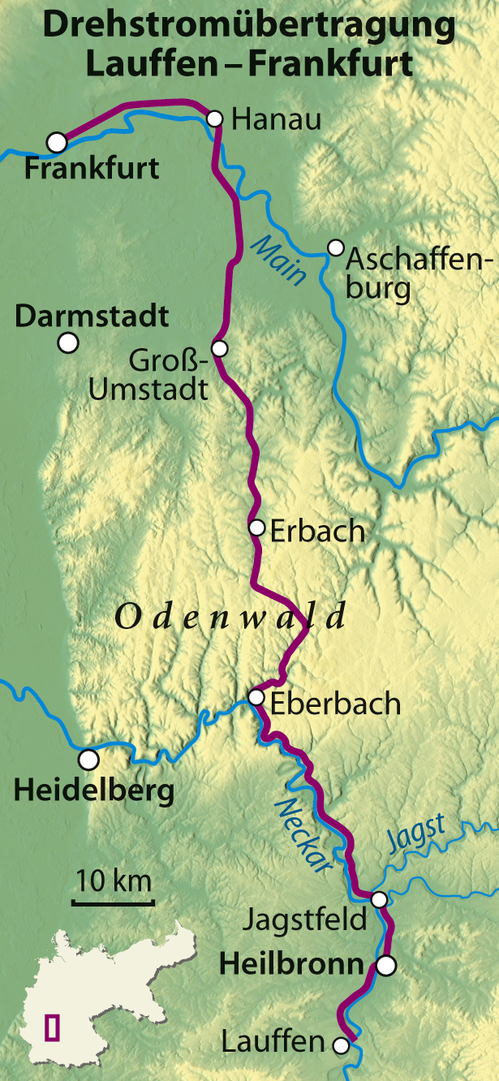}\label{firsta}}\qquad%\hfill
     \subfloat[]{\includegraphics[height=0.35\textwidth]{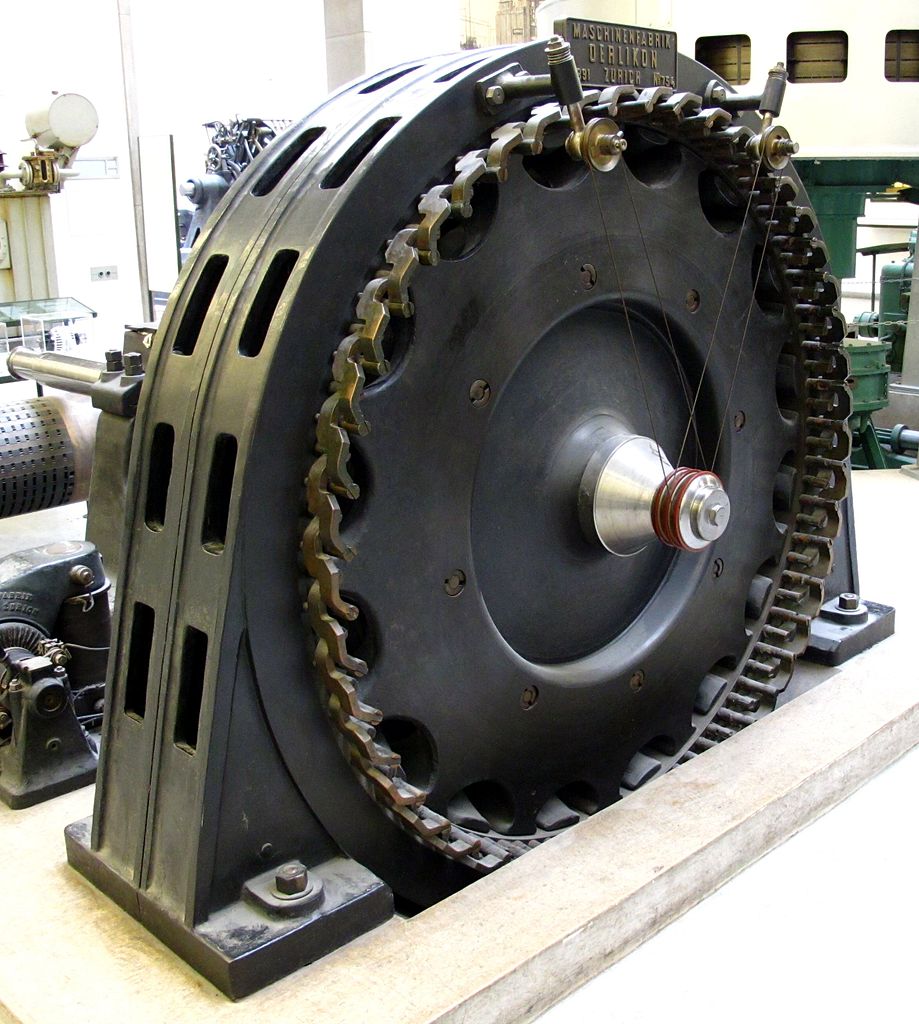}\label{aeg}}\qquad%\hfill
    \subfloat[]{\includegraphics[height=0.3\textwidth]{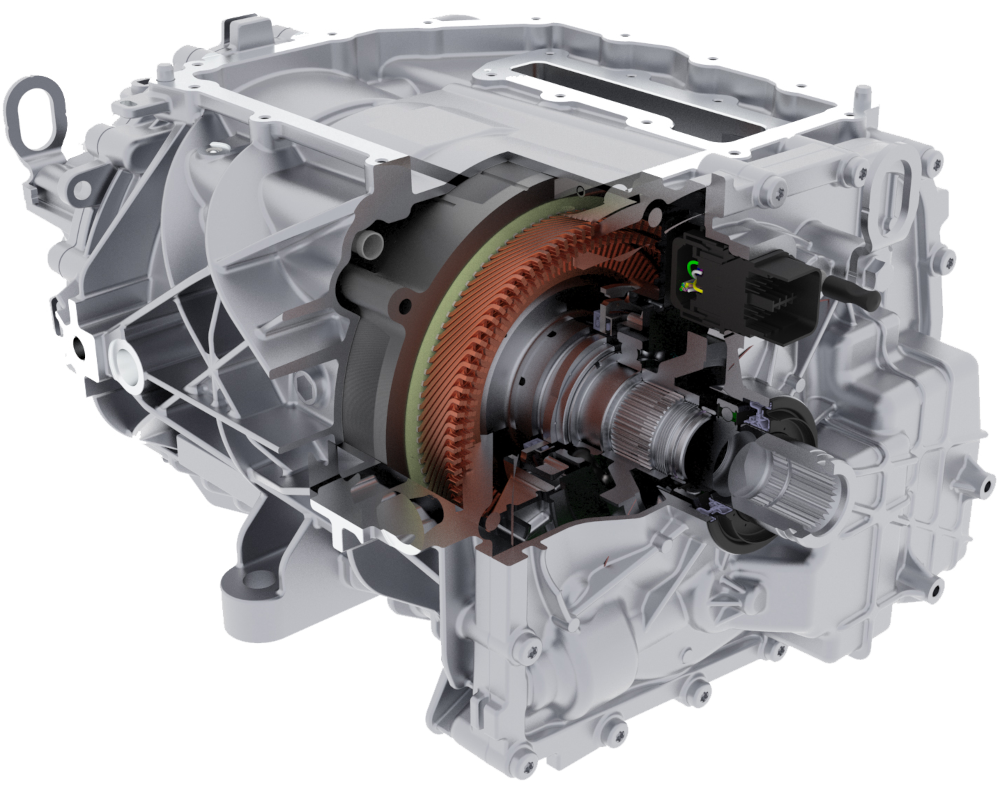}\label{borgwarner}}

    \caption{130 years of improvement of conductor bars in electrical machines. (a) Map of the Lauffen--Frankfurt transmission line in 1891 (now v.\ Miller's Deutsches Museum Munich). (b) Oerlikon Company 223~kVA 55~V three-phase generator in Lauffen 1891. (c) BorgWarner (former Delco Remy) 400~kW 800~V high-voltage hairpin (HVH) 320 Electric Motor 2020  BorgWarner }\label{first}
}
\end{figure*}

A pioneering application of bar winding machines can be found in the Lauffen--Frankfurt three-phase power transmission system displayed in Fig.~\ref{first}. In 1891, this demonstrator (fostered by dc transmission pioneer O.\ von Miller) was able to transmit energy to the International Electro-Technical Exhibition in Frankfurt (chaired by O.\ von Miller) from a hydro generator located over 175~km away at the local cement works in Lauffen, transformed to medium voltage through three-leg transformers \cite{US422746A,esej_19920050,8944322}. As widely known, this demonstration cleared many doubts related to the practicality and feasibility of long-distance three-phase ac electric power transmission with a 75\% efficiency, besides being one of the pivotal events marking the end of the often cited War of the Currents between ac and dc power \cite{HaselwanderPat,CH4160,CH3455,Thompson,esej_19920050,8944322,4295047}.\footnote{With the pioneering work of C.\ Brown, D.\ v.\ Dobrovolsky, and importantly also F.\ A.\ Haselwander, the newly founded General Electric, formed after a major merger of T.\ A.\ Edison's  companies with Thomson--Houston Electric Company at the end of Edison's tenure, adopted this new three-phase ac system under E.\ Thomson's lead in a generator designed by C.\ P.\ Steinmetz already for their bid for the Niagara Falls Adams Power Plant. The winning quote from competing Westinghouse Co.\ used bar windings, though with the older two-phase system. A phase-shifting transformer re-shaped the two-phase power to three phases for transmission to Buffalo (named Scott connection or Scott-T transformer later after Charles F.\ Scott, Westinghouse engineer and AIEE president), where a part of it was changed back to two-phase electricity for distribution, though  \cite{1269626,ScottTransformerA,ScottTransformerB}. In the Mill Creek Hydroplant, Redlands, California, less prominent but a few years earlier (planned 1891, commissioned 1893), however, General Electric successfully installed a three-phase system with generator designs by A.\ Decker. This is considered the first commercial three-phase system in the US, developed and installed almost immediately after the presentation of the Lauffen--Frankfurt demonstrator. The generators, furthermore demonstrated synchronization means to power up and connect each of the initially two generators to an already energized system.}

This three-phase claw-pole synchronous generator proposed by Charles E.\ Brown and Michael v.\ Dolivo-Dobrowolsky was rated for 300~hp and 150~rpm, with 32 poles. Similar to modern three-phase hydro-power generators, the high number of poles was associated with a relatively low rotational speed, a high active diameter of 1,752 mm, and rather short active length of just 380 mm, which also required short end-turns enabled by the bars with connecting clamps to limit the leakage flux. The terminal voltage was in the 50~V to 65~V ac range while the current reached 1400~A, hence justifying a high-current bar design reaching a remarkable reported efficiency of 95.6\% \cite{ETZVienna:1892, Mueller:2001, esej_19920050}. The stator consisted of 96 solid round copper bars with a diameter of 29~mm each in separate circular closed slots and insulated with asbestos sheets. These bars were arranged in a simple wave winding with one bar per pole per phase, where the bars of each phase could be connected in series \cite{esej_19920050}. The wave winding turned out a cunning design to relay the stator current around the large active circumference and collect the phases at the terminals on top, right next to a clamp forming the neutral point (see Fig.\ \ref{first}\subref{aeg}). Accordingly, the design did not require any jumper wires to bridge gaps or connect irregularities, but the three phases and their neutral point terminated right next to each other, using only identical radially c- and axially s-shaped clamps  with a pitch of three to form the coils. Marking the commencement of three-phase systems, it implements the simplest possible winding scheme---single-layer, full-pitch, and integer-slot---which was already distributed but not yet optimized for a smooth air-gap flux and perfect sinusoidal output. The end-turns present a solution with two radial and two axial levels for guiding the bars around each other. In modern terms, such bars that require interconnections through clamps, brazing, or welding on both sides are often called I-pins \cite{US1789128}. When compared to a modern bar-wound machine, displayed in Fig.\ \ref{first}\subref{borgwarner}, the influence of such an early example is still remarkable, despite motors at present featuring about twice the power at a fraction of the size. Most of the gains stem from improved cooling and higher speed, which the improvement of the bar-winding technology over the years enabled.

\begin{figure*}[tb]
\centering{
 \includegraphics[width=0.7\textwidth, angle=90]{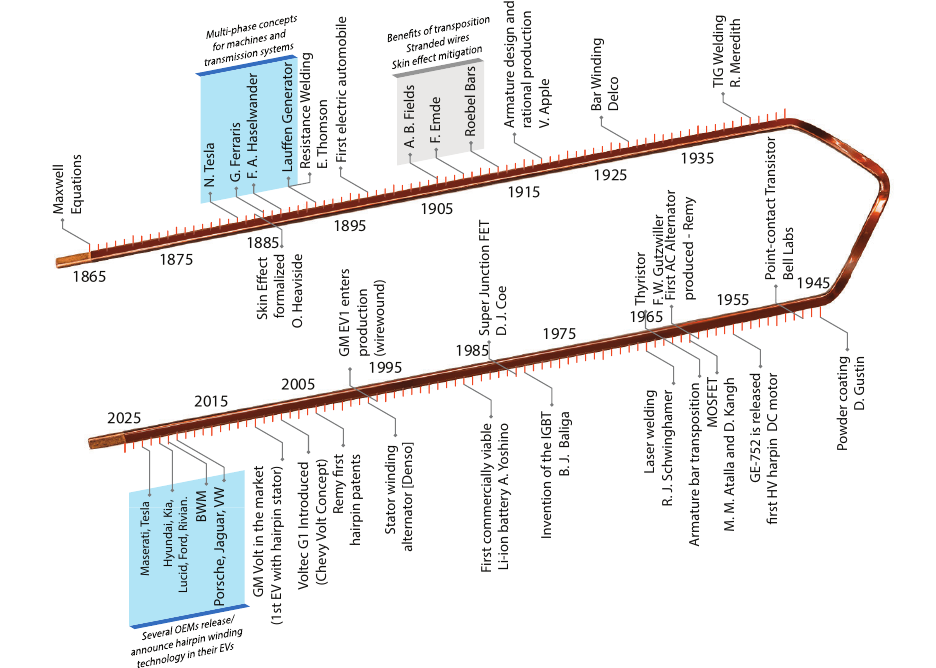}
 \caption{Timeline showing the historical development of hairpin winding technology besides some other breakthroughs related to hairpin manufacturing, such as resistance, tungsten-inert-gas, and laser welding, variable speed drives (power transistors), and electric vehicles \cite{ETZVienna:1892, 6487583, ResistanceWelding, TIGWelding, Thyristor, SuperjunctionFET, IGBT, PowderCoating}.}\label{timeline}
}
\end{figure*}

%\textbf{mention how the need for higher speeds in turbine synchronous generators }

\begin{figure*}[!t]
    \centering
    \includegraphics[width=0.75\textwidth]{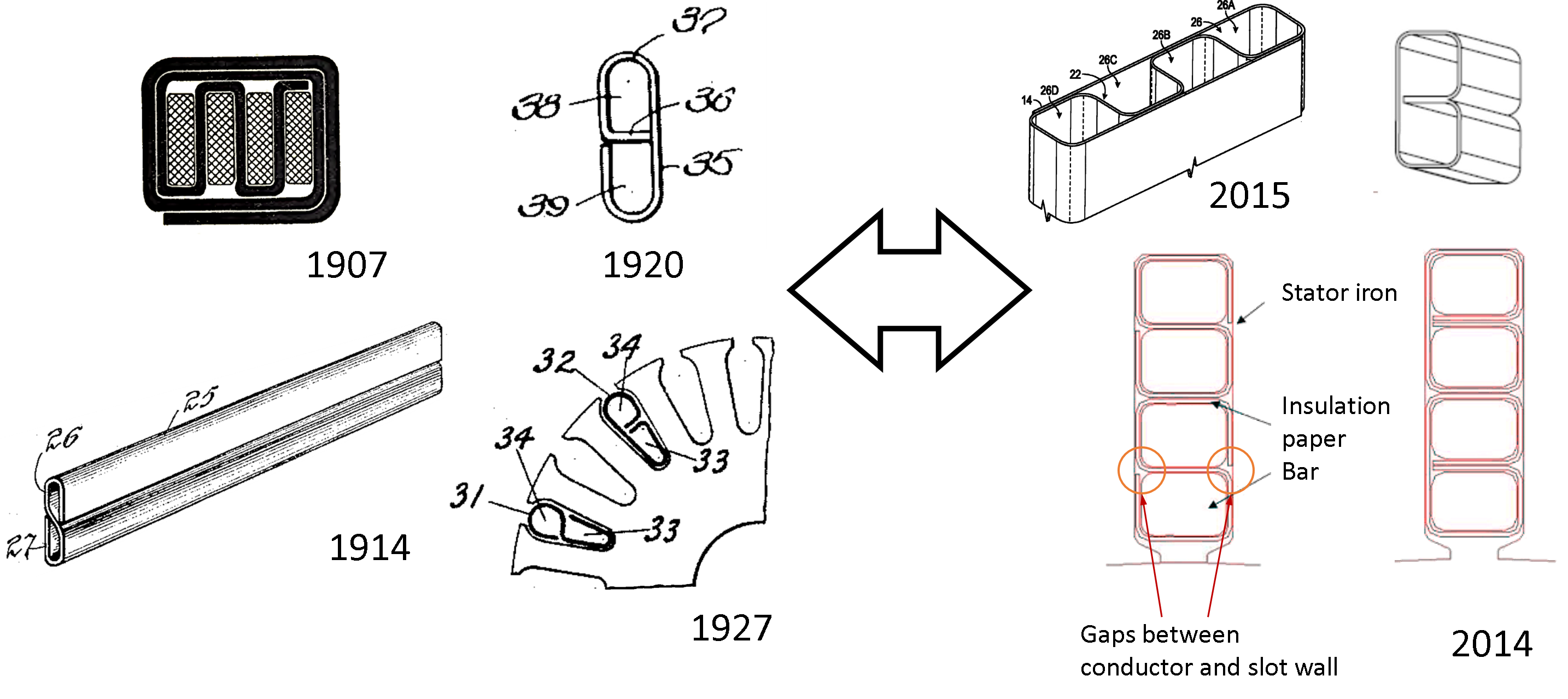}
    \caption{Bar windings and hairpins allow a well-defined and thermally improved contact between conductors as major heat source themselves and to the slot walls while insulation paper can be used for both but should be as thin as possible and leave no gaps for high voltages. Innovations presented in the ongoing wave of hairpin designs and subject of patent applications \cite{US20150263578,Hawkins2014,Rahman2016}, however, turn out to have substantially close historical, though likely forgotten counterparts \cite{US1555931, US1742190, Hobart1907, US1224518}. B-shaped slot liners that meander around all conductors as recently promoted, offer gap-free insulation with sufficient creepage and no larger overlaps but were suggested already by V.\ Apple and others about a century  earlier with similar arguments and even drawings \cite{US1742190}.}
    \label{fig:insulationpaperhist}
\end{figure*}

%The claw-pole synchronous machine reached a power level of 223 kVA at a frequency of 40 Hz (150 rpm). The low voltage of approximately 55 V despite the high power entailed high phase currents in the machine of 1400 A and 
 %The project was headed by Michail v. Dolivo-Dobrowolsky, lead engineer at AEG, and by Oskar v. Miller, head of the International Electro-Technical Exhibition of 1891, an early pioneer of High-Voltage Direct Current Transmission (HVdc), and Director of AEG, the former German Licensee of the Edison Co.\cite{Miller-Biographie}, and implemented by Maschinenfabrik Oerlikon \cite{Thompson,Neidhofer1992:1992, Mueller:2001}. %\textcolor{Red}{At the same exhibition, several further bar-wound machines were presented, among them, for instance, a machine}

{{At this international exhibition, several further bar-wound machines were presented, among them, for instance, a three-phase ventilator motor with solid copper bars provided by the same AEG \cite[pp.~17f.]{ETZVienna:1892}. Bars, themselves straight, were linked by interconnection pieces: c-shaped in axial direction, s-shaped in radial direction so as to pass the skipped bars and their interconnection pieces without contact; welded on both ends.} According to reports, the machine reached efficiencies of 83.0\,\% at 60~hp, and even 93.5\,\% when generating 190~hp \cite{esej_19920050}. }

Many of the contemporary energy generation and transmission projects used some form of bar winding, showing an important increase in the voltage range compared to the Lauffen machine. The two-phase Niagara generators with their twelve poles were rated for more than 2,000~V for 250~r/min, while using already two bars per slot \cite[p. 317]{Houston:1896}. In 1895, the three-phase 800~V Folsom Power House generator designed by C.\ Steinmetz and E.\ Thomson at the same time demonstrated the search for the ideal design of ac machines and the incremental developments on the way from dc.
%Like at least one of the Lauffen machines, also several Folsom generators have survived to this day.
Rather unusual today, the rotor of this in-runner incorporates the ac winding on the rotor and fixed poles on the stator---just as already before the Mill Creek generators from the same designers. The ac winding on the rotor indicates how ac technology to some degree evolved from dc motors, where the armature with the ac current is likewise on the rotor. This design practically just dropped the mechanical commutator, which previously generated the armature ac out of terminal dc current. On the other hand, the winding scheme was already improved compared to the Lauffen machine with two bars per slot and formed the rotating field, whereas the stator incorporated 24 static excitation poles \cite{US522241}. Slip rings (but without commutation) fed the current to the armature winding on the rotor. As this early ac machine was practically a dc motor with slip rings instead of the commutator, it could re-use many of the existing dc components. The connections at the end-turns were brazed according to the state of the art at the time.

\begin{figure*}[!t]
    \centering
    \includegraphics[width=0.8\textwidth]{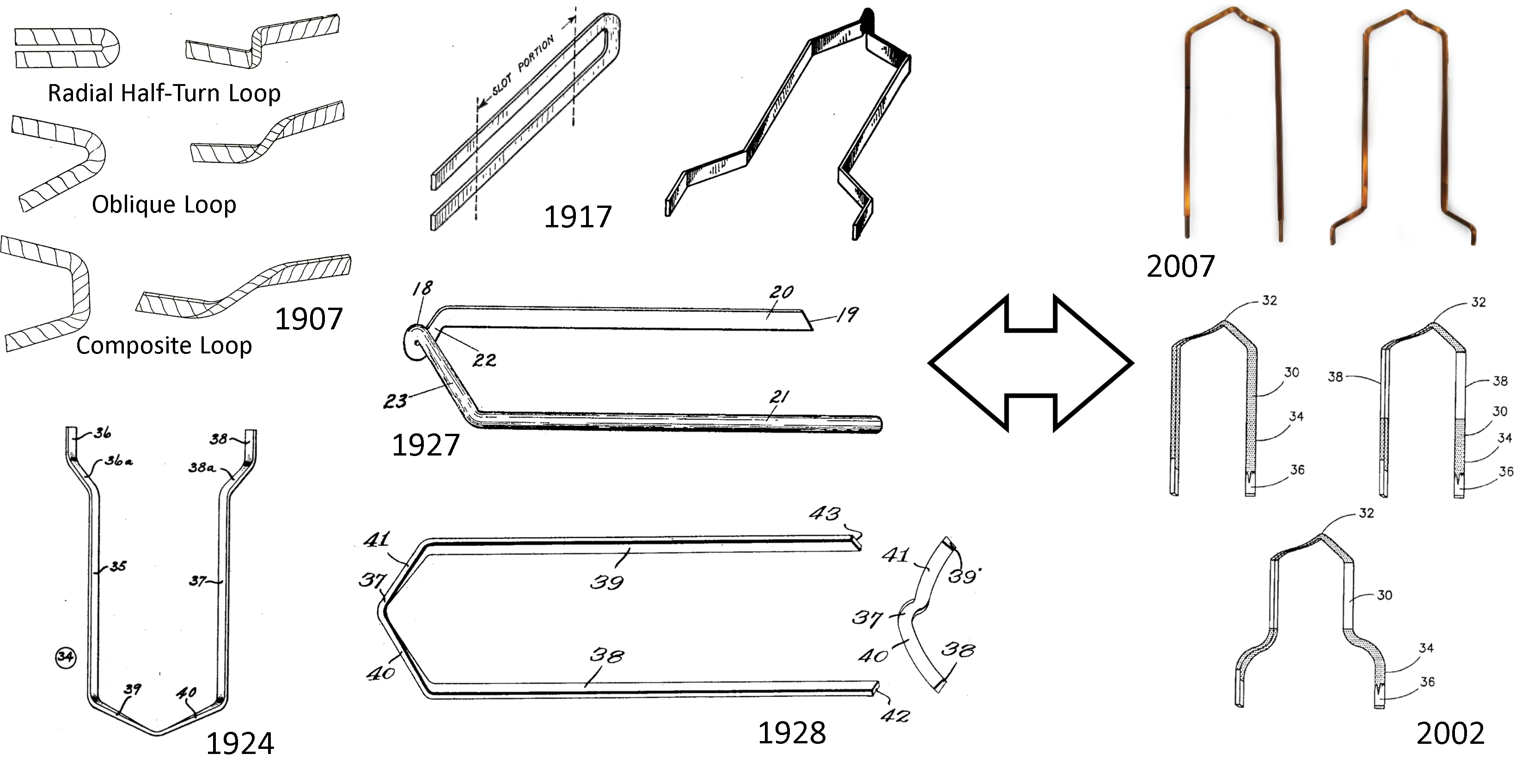}
     \caption{Historical hairpins \cite{Hobart1907, Hawkins1917, US1808749} vs.\ modern ones \cite{US7805825, US6894417}. While hairpins had been out of fashion for a long time before their rediscovery, some features survived in form windings for large machinery \cite{}.}
    \label{fig:hairpinshapeshist}
\end{figure*}

Another important application were electroplating generators, where bar windings clearly outperformed wire-wound designs. The bars could provide ample cross sections for the large currents, while the voltage typically was comparably low, often in the single-digit Volt range. Prominent manufacturers were Soci\'et\'e  des Machines Magn\'eto-\'Electriques Gramme, which were known for its machines with low ripple content (Gramme machines), and Siemens  \cite{Siemens1881, Siemens1891, GlaserDeCewKrohnHiggs:1884, siemens1883science, higgs1878some}. % is Siemens1891 == MartinBook:1992?? Discusses Kupferstab (Suche) on Page 451
These dc alternators were optimized for high current and heating tolerance, later also for a stable voltage. Asbestos served to insulate the bars from the slot walls, which offers high temperature stability if its hygroscopic trait can be managed \cite{Arnold:1902,Turner:1906,6537640}. Even the insulation of bars with surface coating, which is an essential component of modern hairpin windings, was explored with shellac and hard rubber (ebonite) \cite{Arnold:1902}.
%\cite{Arnold:1903}.

%\textcolor{red}
{Figure \ref{timeline} sorts and highlights some of the most relevant discoveries around bar and segmented windings throughout the years. The following sections will discuss the most prominent ones. Some additional flashlights related to electric vehicle technology in the timeline figure contextualize historical development trends.}

\begin{figure*}[!t]
    \centering
    \includegraphics[width=0.7\textwidth]{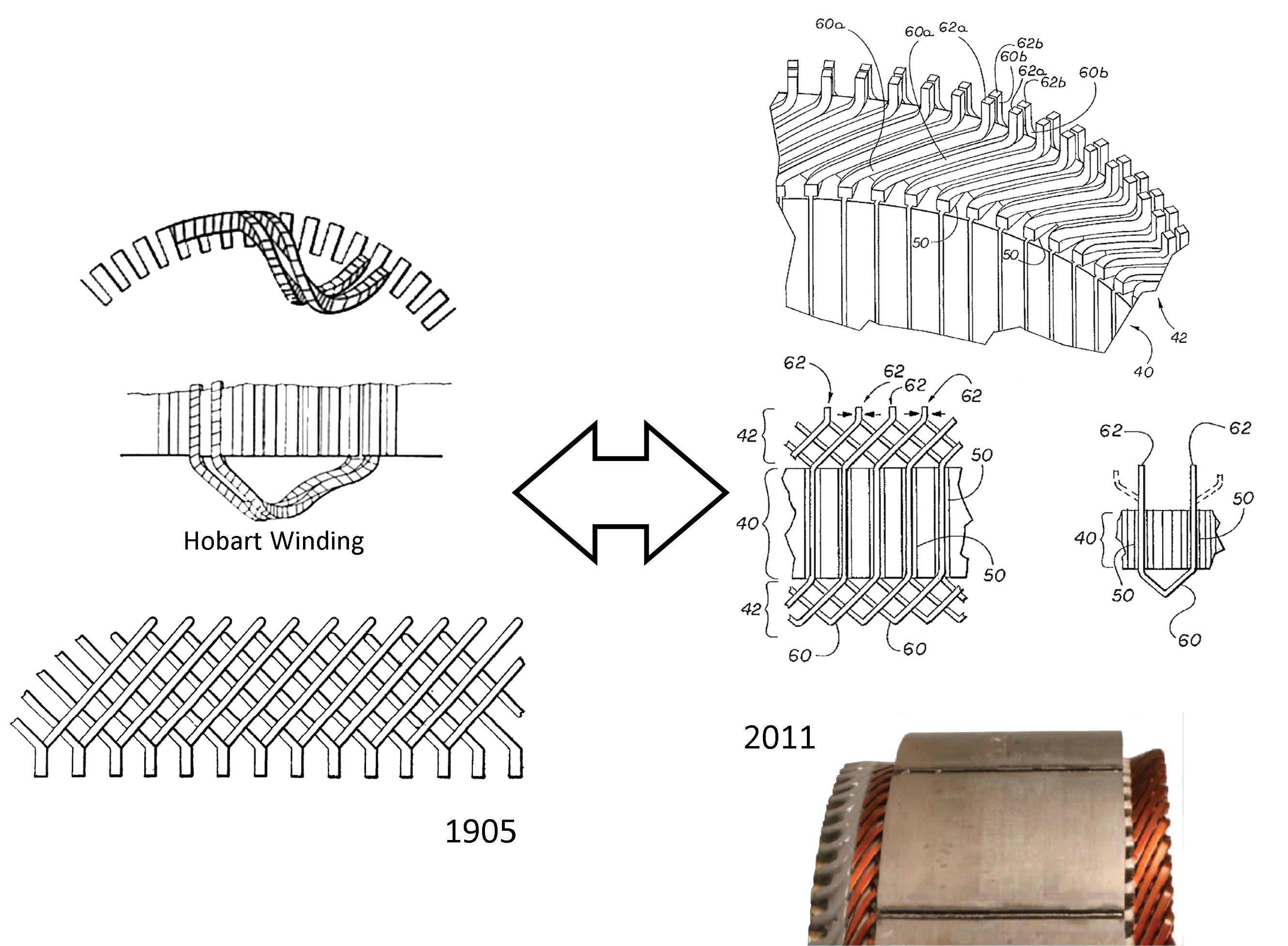}
    \caption{A major art in hairpins is a smart pattern at the end turns to reduce the size of the overhang and guide the conductors around each other. While M.\ v.\ Dolivo-Dobrovolski had already found the  fundamental braided three-phase pattern using two radial positions for ascending and descending bars, the more modern shape using bent wires evolved around the turn of the century \cite{Hobart1905, US8499438}.}
    \label{fig:endturnstructurehist}
\end{figure*}

\begin{figure*}[!t]
    \centering
    \includegraphics[width=0.85\textwidth]{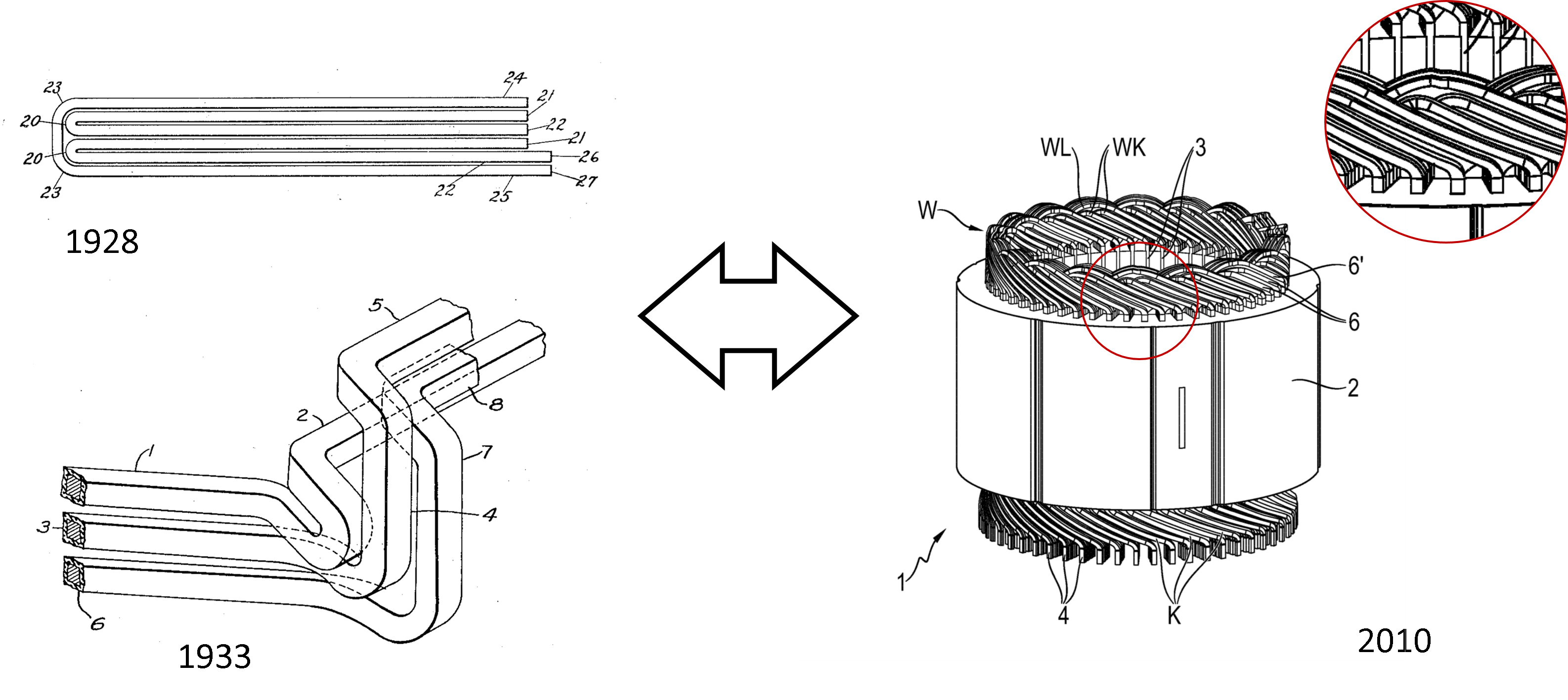}
    \caption{Of major importance in hairpin and bar windings is the control and suppression of high-frequency losses. Recently a number of transposition techniques have been proposed, which change the position of bars within slots \cite{US20220360128, DE102018125829}. However, the importance of the problem as well as solutions were already developed long time ago \cite{US1826295A, US2085099}. The prior art provides a rich knowledge source for such aspects.}
    \label{fig:transpositionshapes}
\end{figure*}
 
\begin{figure*}[!t]
    \centering
    \includegraphics[width=0.75\textwidth]{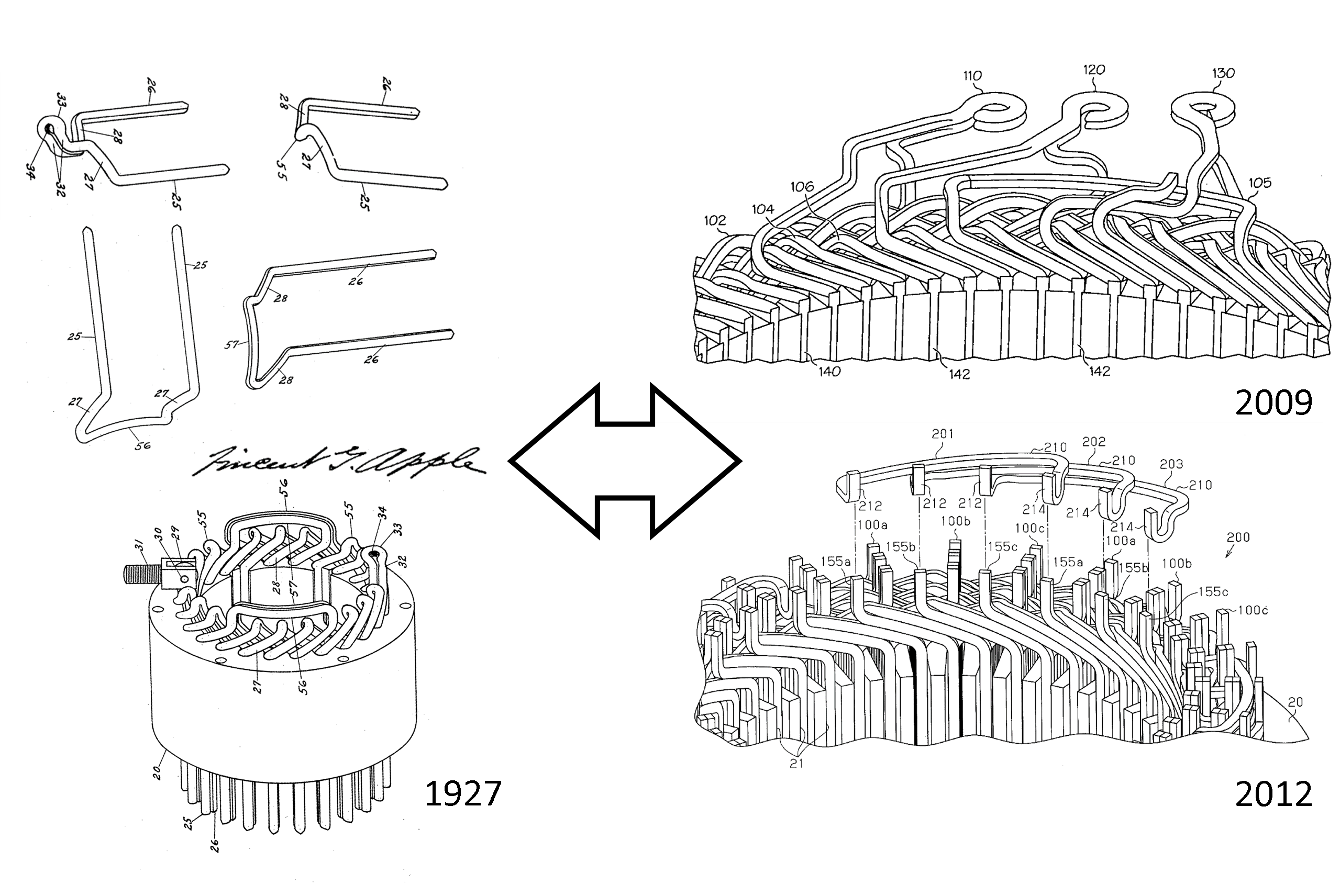}
    \caption{Jumper wires and collectors with more complicated bending section of various kinds can form terminals, neutral points or delta connections, as well as wider bar transpositions \cite{US20090140596, US20120200191}. Earlier inventors during the first hairpin wave already suggested many of such techniques, among them also V.\ Apple \cite{US1822261}.}
    \label{fig:jumperhist}
\end{figure*}

\begin{figure*}[!t]
    \centering
    \includegraphics[width=0.85\textwidth]{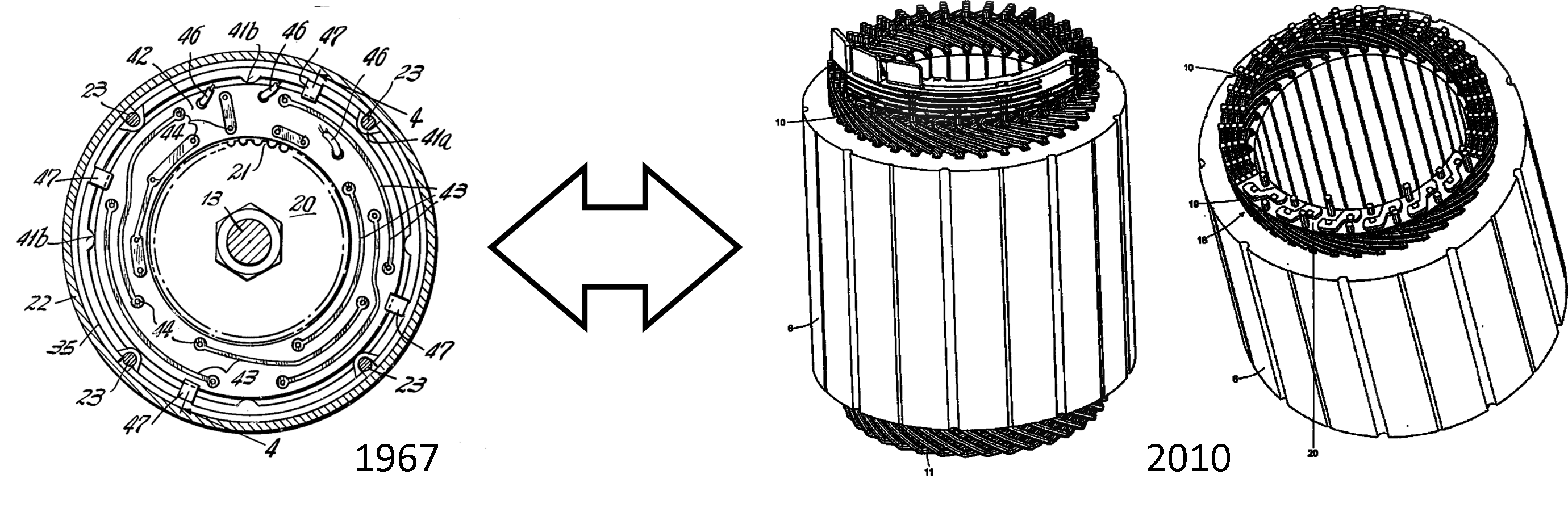}
    \caption{A rather compact way for connecting individual phases and form terminals are (layered) circuit or connection boards, which have become the hallmark of some manufacturers \cite{EP2437378}. As this solution is attractive, precursors exist and survived to this day in automotive 12~V alternators \cite{US3495109, US3041484}.}
    \label{fig:layeredTerminals}
\end{figure*}

\begin{figure*}[!t]
    \centering
    \includegraphics[width=0.9\textwidth]{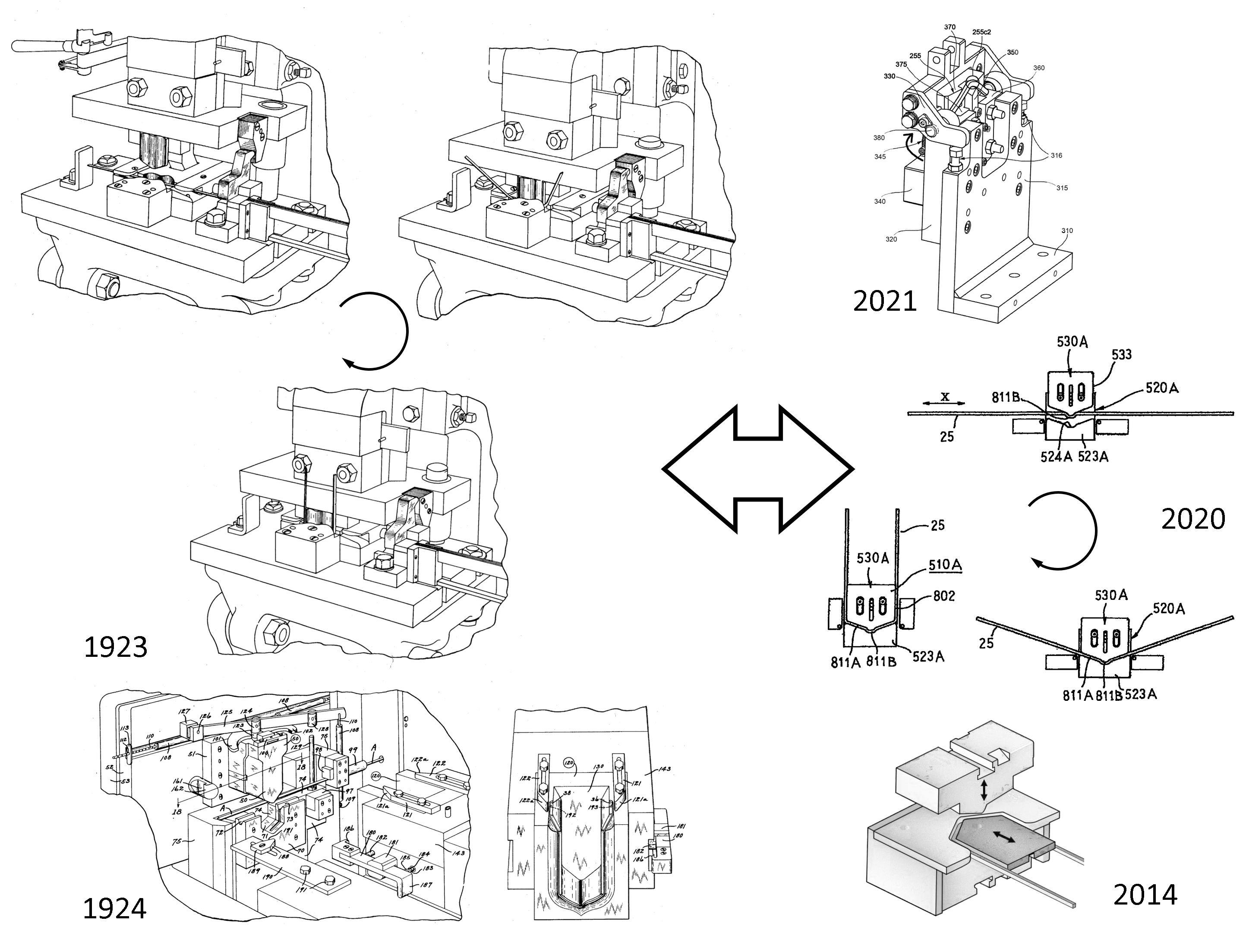}
    \caption{Hairpin windings offer a highly economical fully automated manufacturing process with automated bending, insertion, and particularly interconnection process. Particularly bending was already professionalized a century ago\cite{US1721636, US1721810A}. The machinery looks very modern and close to contemporary devices \cite{EP3811504,US10594195,Hawkins2014}.}
    \label{fig:bendingmachinesA}
\end{figure*}

\begin{figure*}[!t]
    \centering
    \includegraphics[width=\textwidth]{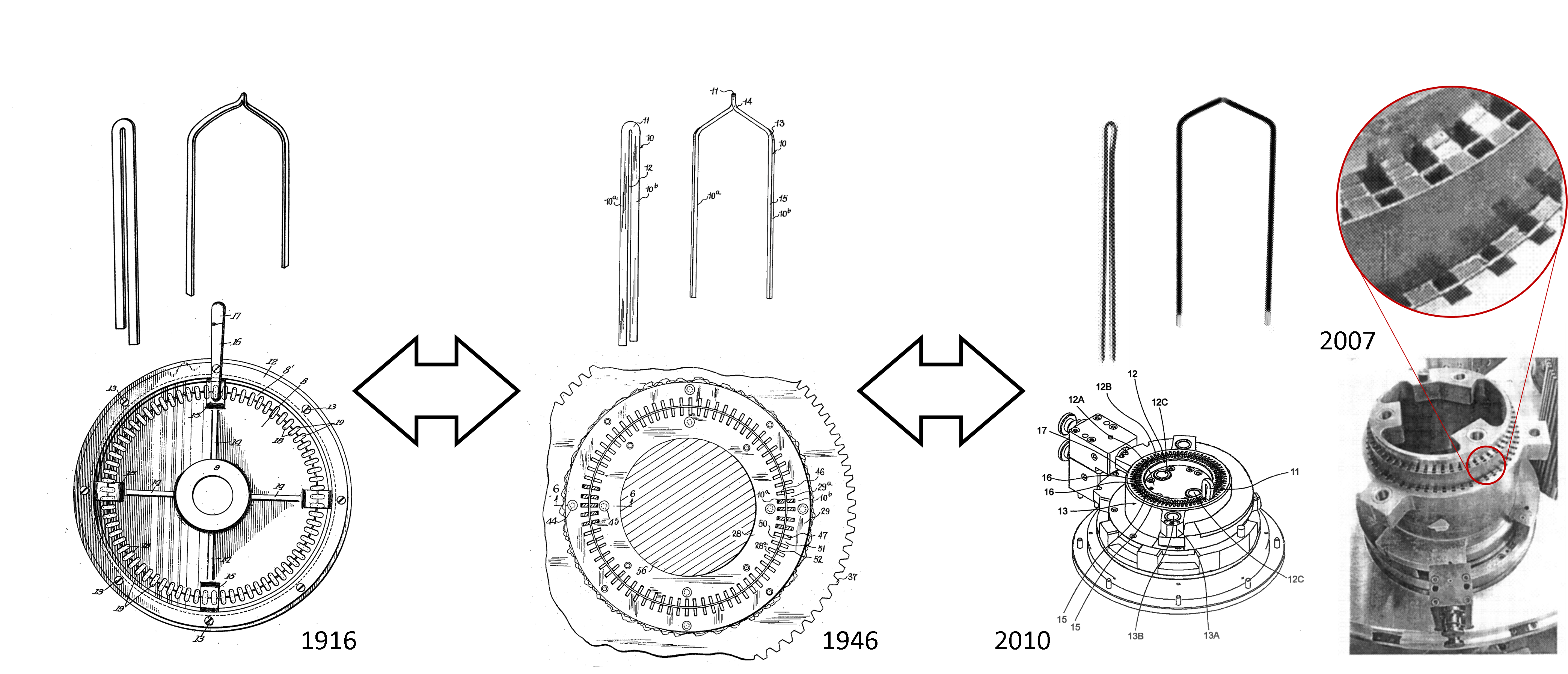}
    \caption{As an alternative to hairpin forming with molds in presses (Fig.~\ref{fig:bendingmachinesA}), hairpins are routinely formed through inserting closed hairpins into harnesses where the two legs are pulled apart when the respective discs that hold them rotate relative to each other \cite{US8561447, US7805825}. This technique may have less control over the end-turn shape but intrinsically adjusts the angle between the corresponding sides of the legs according to the slot walls when the radius of the discs are matched to the slots of the machine that should receive those pins. The idea is almost as old as hairpin windings and figures in patent applications appear dauntingly similar \cite{US1238959A,US1555931,US2476743}.}
    \label{fig:bendingmachinesB}
\end{figure*}

\begin{figure*}[!t]
    \centering
    \includegraphics[width=0.7\textwidth]{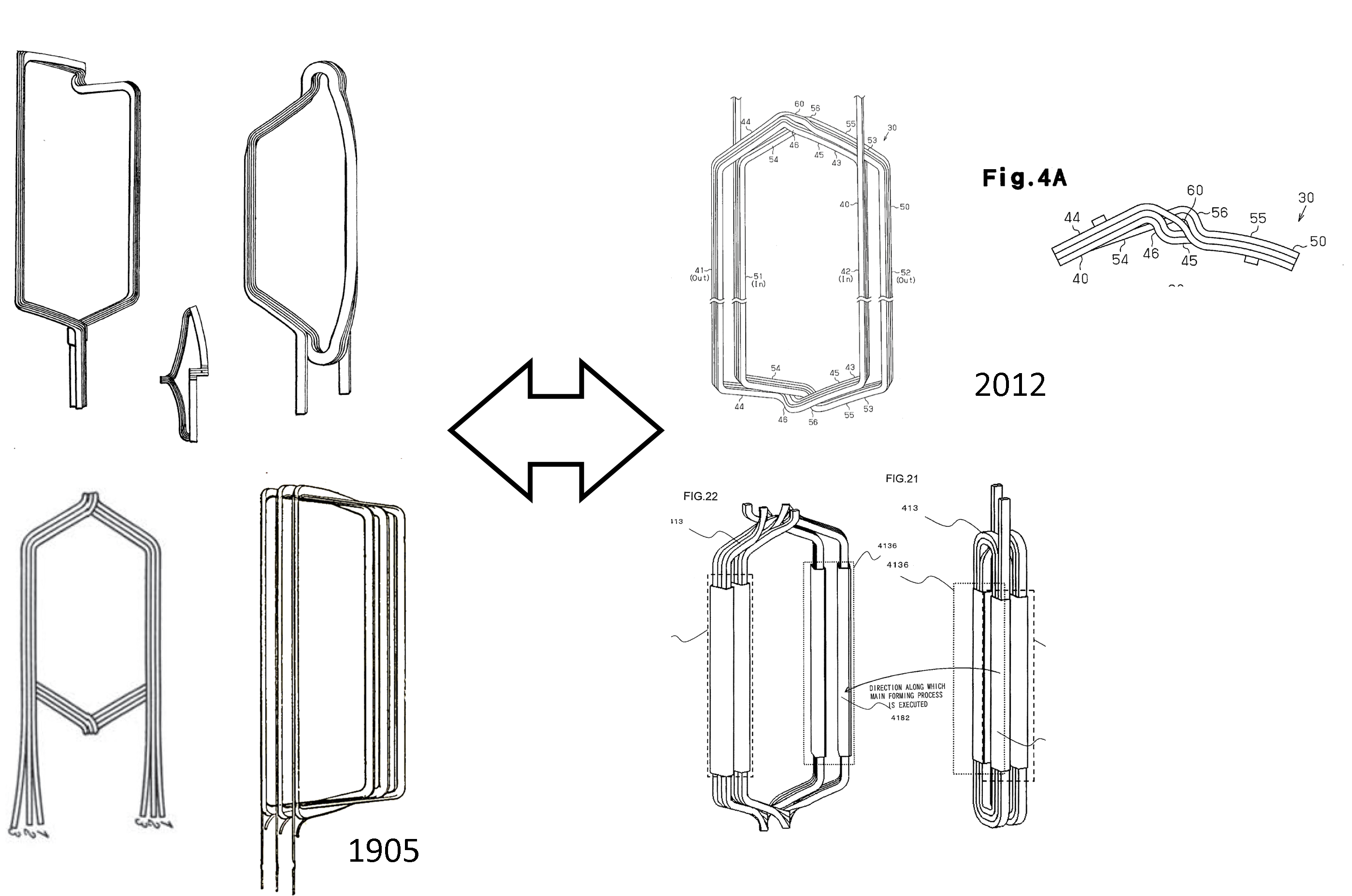}
    \caption{Longer segments than U-shaped hairpins are particularly used in lap windings and sometimes called \textit{diamond coils} \cite{US20120200191,US8330318}. Those include winding elements that have turns in more than two slots, \emph{e.g.}, for higher winding factors and/or skewed windings \cite{Hobart1905, Hobart1907, US3453468}.}
    \label{fig:diamondcoilshist}
\end{figure*}

\begin{figure*}[!t]
    \centering
    \includegraphics[width=.8\textwidth]{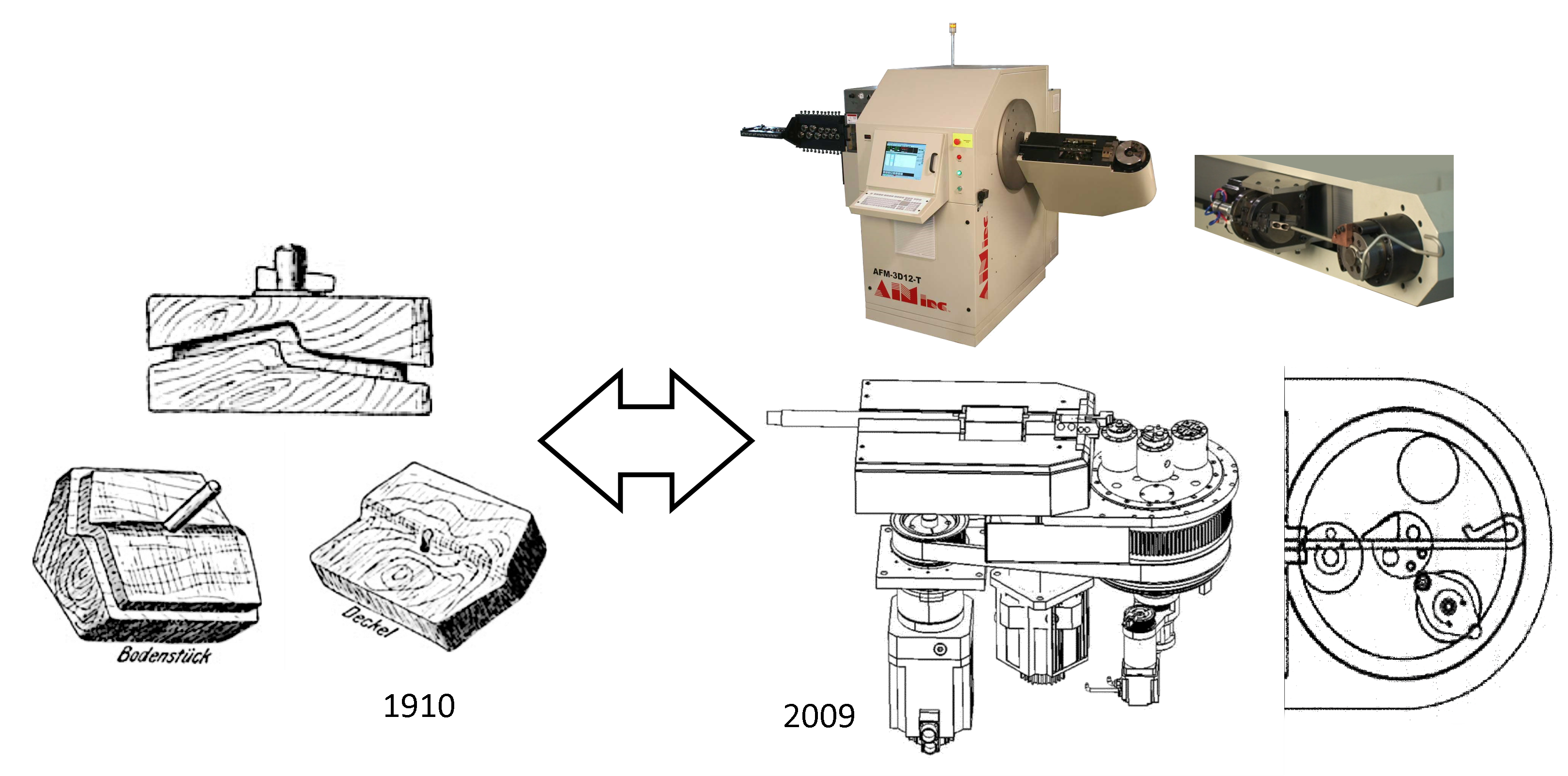}
    \caption{Clear innovations appear to drive prototyping of hairpin and bar windings. Whereas historically formers for the pins that included both the end-turn shape as well as the twisting of the legs' cross section against each other \cite{Krause1910}, CNC bending, such as those from AIM Machines, Inc.,  allow rather fast, even for a larger number of parts or small-scale manufacturing \cite{EP2514536,US10792727}.}
    \label{fig:manufacturingCNC}
\end{figure*}

% \begin{figure}[b!]
% \centering{
% \subfloat[]{\includegraphics[height=0.25\textwidth]{figures/cut_bar.png}\label{bar_w}}\hfill
% \subfloat[]{\includegraphics[height=0.25\textwidth]{figures/cut_round.png}\label{round_w}}

% \caption{Should we include an introductory figure? it is hard to follow, specially the geometry related features. }\label{windings}
% }
% \end{figure}

\section{Form-wound coils as next of kin and mutual inspiration} %parallel development
\begin{comment}
\textcolor{red}{RLF: This section is hard to follow. Moreover, is in base a mechanical concept for the interconnection. The main problem is the next section (IV AC losses) is cover in the surface and I feel like that is presented like a hard problem and, the novel solutions and implementations, they are not fully considered this aspect (I mean when I read the document, specially this section).}
\end{comment}
{Form-wound coils as compacted coil segments made of multiple wires, wire turns, or other conductors packed together to act as one are well-established for large machines and obviously create a link between hairpin or more generally bar windings and wire-wound machines \cite{US561636}.
{Form windings have had substantial influence on the development of the dominant basic machine topologies, and techniques for cost-efficient manufacturing consolidated in the early 20th century. Already in the 19th century, form-wound coils provided means for fast shaping of coils from round wires as well as copper bars or strands and electrically insulating them from the slots before their installation in the machine.
In contrast to the closed slots of the Lauffen generator, half-open and open slots dominate form-wound machines as the windings are typically inserted from the bore. The openings are then often closed with slot wedges.
%Moreover, these coils consisted of round wires for smaller drives, while for medium and large drives typically used copper bars or straps.
The major challenges are the end turns and their shape in order connect the active, in-slot portions of different slots, \emph{i.e.}, how the winding is guided from one slot to another and how the various strands are interwoven to minimize size, copper use, and end-turn inductance. Various forms of such end-turn patterns have evolved; established shapes include rectangular, round, and diamond ends, while diamond end turns shaped as involutes either radially towards the shaft or nowadays more frequently in axial direction has become the dominant design \cite{US596136A,US744680A,Gray:1913,Krause1910}.
%or \textbf{STIRNWICKLUNG}
%influences machine design to this day as shown in Figs.\ \ref{fig6} and \ref{smz} \cite{US596136A,US744680A}.
Other still present developments include half-turn loops at the end turns to form a wide variety of coil shapes without bulges \cite{CH21731A,US561636A,Hobart1907}.}

Form and bar windings share large similarities and have innovated each other particularly with respect to end-turn patterns.
The end-turn patterns that dominate in modern hairpin machines follow the Lauffen approach for guiding bars around each other with typically two modifications (see Figs.\ \ref{fig:hairpinshapeshist} and \ref{fig:endturnstructurehist}): First, the pattern is, in stark contrast to the one bar per slot in Lauffen, extended to two bars which alternate between two radial positions in the slots and have their slot-to-slot connections on opposite sides of the machine, generating advantageous clearance in the end-turns for welding and direct cooling. The bar closer to the air gap leaves the slot on the front side to enter the next slot after the respective coil span at the radially outer bar position, \emph{i.e.}, farther away from the air gap. The two or more bars per slot (typically multiples of two, following the alternating pattern within each pair) allow short-pitched double-layer windings, which was found to not only improve the air-gap flux smoothness at the cost of a marginally lower winding factor---the usual reason for short pitching---but also a potent way for reducing ac effects in the large cross-section conductors \cite{US6894417}.
For two such bar positions and a coil span of $s$, the pattern needs to guide $2s$ bars around each other and does that---following the Lauffen design---by forming several radial and axial layers in the end turns (Fig.\ \ref{fig:endturnstructurehist}). Most commonly, end-turns reflect the two radial positions that nowadays already exist in the slots and establish $s$ axial layers \cite{RahmanSAEI}.  Second, the overhang follows a triangular or diamond shape, at the end of which the bar changes its radial position before it dives back down. This cusp is sometimes called \textit{knuckle} and can be on the bent as well as on the jointed side (see Reference Numbers 18, 32, and 37 in Fig.\ \ref{fig:hairpinshapeshist}) \cite{Liwschitz:1950}.

Modern hairpin windings typically have a higher number of rows as a multiple of two (commonly 4, 6, 8, 12) to adopt the fundamental interwoven pair design. Short pitching, e.g., through shifting the rows against each other, and overall more complicated winding schemes may not always work with just one type of pins and require so-called jumpers. Jumpers are pins that deviate from the typical pitch or pattern, jump from one row to another, or connect more distant pin ends. Concepts for simple jumpers that blend in well, however, have been developed early on (Fig.\ \ref{fig:jumperhist}). Such jumpers can also be used to form the phase terminals. The use of layered jumpers and terminals that remind of printed circuit boards and allow axial and radial bypassing appear smart and compact and are therefore used and promoted intensively. However, such concepts have also been suggested already earlier (Fig.\ \ref{fig:layeredTerminals}).

%\begin{figure*}\begin{center}
%    \includegraphics[width=0.75\textwidth]{figures/knuckle.png}
%\end{center}
%    
%\end{figure*}

Although both wave and lap windings are possible, hairpins nowadays predominantly use wave windings as already the Lauffen generator did. Bar windings with longer segments than I-pins and hairpins are also available, typically as lap windings, and resemble what was once developed as so-called spread-diamond coils (Fig.\ \ref{fig:diamondcoilshist}) \cite{US20100001609A1, Ishigami:2014, US8330318, Krause1910, Liwschitz:1950}.

The change of the radial position of bars (\emph{e.g.}, within a pair) from slot to slot practically pairs two Lauffen windings, interweaves them, and generates a regular pair-wise transposition, which alleviates high-frequency ac effects to some degree. This pattern has been well established for more than 100 years \cite{Arnold:1902}. Transposition patterns for more bars have been devised already early on and currently re-discovered or invented to guide a phase's current through ideally all possible positions inside the slots to balance different magnetic field conditions (Fig.\ \ref{fig:transpositionshapes}). For that purpose, pins can for example reach over a pair forming a kind of arc. Even with more than two conductors per slot and potential transposition patterns, most modern hairpin machines still group their bars in pairs, each of which follows this long-established design.
}

By 1910, form- and bar-wound coils had become standard for dc machines \cite{Arnold:1902,Krause1910}. For ac machines, they were slowly gaining more importance. The slower development had various reasons. First, the typically higher voltages and therefore lower currents of ac machines did not necessarily justify the higher slot fill; second, ac machines were still underdeveloped compared to dc machinery \cite{Arnold1908,esej_19920050}. By that time, also three-phase induction machines were tested with solid bars for the stator winding to increase the phase current. However, the increased complexity of the interconnection led to typically not more than one to four bars per slot in one or two layers with both lap windings for higher voltage and wave windings for lower voltage machines \cite{Arnold1908}. Insulation with closed tubes of oil paper, oil canvas, calico cloth, varnish, and mica already enabled phase voltages of several kilovolts \cite{Arnold1908,Turner:1906}. Early enamel coatings had thicknesses of approximately 0.1~mm and allowed denser packing than cloth or bandages \cite{Arnold1908}.

% reference unknown AltesBuchzuQuerschnitten
\section{Discovery of ac losses and associated retreat of bar windings to niche applications}

Considering that early bar-wound machines used relatively large-cross-section conductor bars, an important constraint was and still is additional loss due to the skin effect, proximity effect, and copper eddy currents. Although J.\ C.\ Maxwell had already discovered these effects, he and his contemporaries may have underestimated them or otherwise not have perceived their practical implication for some time \cite[p. 487ff.]{ARussell:1904}. The topic of ac effects and their knowledge and appreciation by the relevant communities in engineering appears more complicated and not conclusively black and white.
It was previously even claimed that the skin effect in slots was not known in the 1890s and had apparently not been fully recognized for their impact before the work of A.\ B.\ Field after the turn of the century \cite{esej_19920050, Field:1905}. Early reports of the various ac effects however must in fact have been clear to the informed scientist, although their impact may have been underestimated. Eddy currents in particular---including those in iron parts and the conductors---were almost certainly known by v.\ Dolivo-Dobrowolsky, Brown, and others given the coverage in the professional literature \cite{US392765A,US382279,US422746A,bruce1889}; the skin effect on solid conductors, then also called Thomson effect after W.\ Thomson,  and means to suppress it in windings, such as stranding, were also explored and discussed at this time in patents and engineering societies \cite{RSMeetingCrookes1891,US645014}.
%
%Although J.\ C.\ Maxwell had already discovered these effects, he may not have recognize their nature entirely \cite[p. 487ff.]{ARussell:1904}.
%Hughes also experienced an increase of resistance of telegraph wires with frequency while exploring self-inductance in the 1880s \cite{Hughes:1886} and knew about the improving effect of stranded wires \cite{Hughes:1887}.
O.\ Heaviside and J.\ W.\ Strutt, a.k.a.\ Lord Rayleigh, studied the skin effect analytically in the 1880s \cite{Heaviside:1884, Rayleigh:1886a, Rayleigh:1886b}, whereas the detrimental effect of the skin effect on ac machines was already identified and clearly communicated in 1889 \cite{bruce1889}, \emph{i.e.}, approximately the time of the design of the Lauffen generator.\footnote{W.\ Preece concludes that ``Sir William Thomson had shown, principally by mathematical reasoning, that if the currents were sent with a high frequency, they had not time to penetrate into the interior of a conductor. With a solid conductor, say between Deptford and London, 1 inch thick, at a high frequency, the current would not enter more than about 3 millimetres inside the surface'' \cite{bruce1889}.} In the following time, E.\ Merritt presented spatial distribution plots, rendering the effect of current concentration on the surface obvious also to non-theoreticians \cite{Merritt:1897}. With A.\ Russell's and A.\ Field's work on skin and eddy-current phenomena, sufficient knowledge about the adverse effect of large cross sections became available in the literature \cite[pp. 487ff.]{ARussell:1904}, \cite{Field:1905}. However, even ten years later, W.\ Rogowski concludes that despite good theoretical and experimental coverage the skin effect is widely not understood and appreciated \cite{Rogowski:1914}. Despite early pheonomenological knowledge by at least part of the community, applicable models and measurement approaches for ac losses in various conductor types and shapes were not available before the early 20th century \cite{Hawkins1917, Dwight:1918, Gilman:1920, Barus:1928, Cockcroft:1929, Haefner:1937}.

The additional losses due to skin effect, proximity effect, as well as circulating currents constrained the use of large conductors in early bar-wound machines altogether. Their discovery revealed the trade-off between ac loss and the cross section of individual bars or parallel strands, as well as the dilemma that in some cases an increased cross section can even reduce the efficiency. Significant advancements led to design rules, and recommendations from that time are still in place \cite{4764567,emde1908einseitige}. Some authors recommended to avoid large cross sections of conductors and prevent the radial stray flux injected by the rotor poles into the slots (instead of the stator teeth) from reaching conductors, \emph{e.g.}, by increasing the distance of the first bar in a slot from the air gap \cite{Arnold1908}. Also, given that the transverse magnetic field in the slot cannot be shielded, tooth saturation should be kept moderate and high copper bars be divided into segments in radial direction; such parallel strands of a conductor phase should not be shorted at the end turns after each segment to suppress circulating currents but rather run through the entire stator to be paralleled at the overall ends only so that imbalances average out \cite{app10217425}.
%Consistently with the understanding of ac losses, bars were primarily recommended for conductor cross sections above approximately 20 -- 25 mm$^2$ \cite{Arnold1908,Say1976}. Such size recommendations made the use of bar conductors rather uncommon for small ac drives in the early 1900s.
Considering ac losses and manufacturing techniques at that time, bars and their high-frequency challenges were primarily recommended for larger currents and cumulative cross sections above 20 -- 25 mm$^2$ as part of segmented form windings so that their use became less common for small ac drives in the early 1900s \cite{Arnold1908,Say1976}. 

Despite the temporary retreat of bar windings into niche applications, the challenge with ac effects inspired important innovations. The experience with litz wire in radio engineering initiated the development of transposition concepts for conductors between two slots, \emph{i.e.}, systematically changing the radial conductor position between slots so that conductors are close to the air gap in one slot and adjacent to the slot ground in another one to average out the flux experienced by each conductor \cite{Say1976,4764567,5060747,5055018,5055556,US1826295A, Fleischmann:1924, Summers:1927}. In addition to short-pitching, transposition of conductors on the way from one slot to another beyond the fundamentally interwoven hairpin pair by shuffling them as described above is a key technique for mitigating high-speed ac losses in bar and hairpin windings to this day \cite{US20180034334, US7242124, US7170211, DE102018125829}.\footnote{Transposition of the conductors within slots at the cost of lower slot fill was suggested early on as the obvious next step, though again with a focus on large drives \cite{US678030A,US680277A}. Similar to later Roebel bars, the presented conductors consisted of strips folded such that each of a pair of adjacent strips fill the free space left by the other one and swap positions in the middle of the slot. The cross section in the thinner section where the strips cross over equals approximately the folded ends \cite{US680277A}. Although this design was considered too expensive for off-the-shelf machines by observers \cite{4764567}, its perfection by Ludwig Roebel \cite{US1144252A} and later modifications are still dominant among large drives \cite{Boldea2016, 4500422, 1325294, 4802267,955524,Pannen:1963, Baodong:1995, Bennington:1970:US3585428, Darrieus:1964:CH431688:FR973566, Darrieus:1961:DE1142408, Darrieus:1960:DE1146184, Macdonald:1971, Tessarolo:2010, Zou:2012, Dordea:2006, Dordea:2009, Dordea:2010, Dexin:2000, Fard:2007}.}

\section{Small Drives in the 20th Century}

\begin{figure*}[t!]
\centering{
%\subfloat[]{\includegraphics[height=0.3\columnwidth]{figures/Cadillac30_advert1912_polished_cart.png}\label{CadillacAd}}\\

\subfloat[]{\includegraphics[height=0.6\columnwidth]{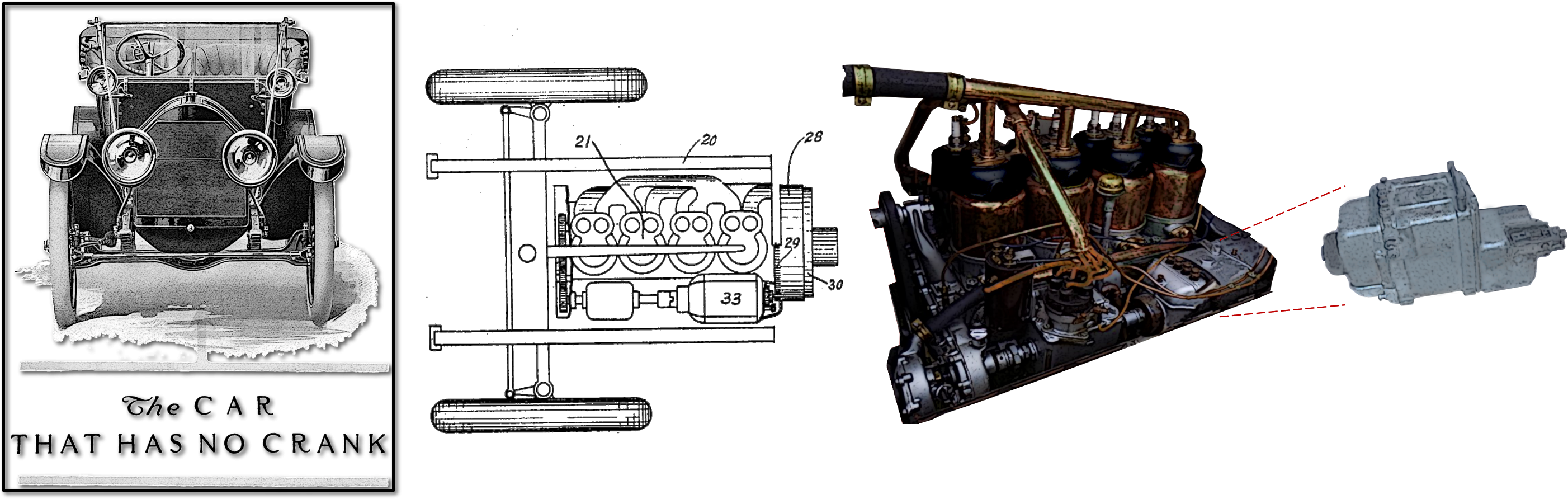}\label{CadillacCollage}}\\

\subfloat[]{\includegraphics[height=0.23\columnwidth]{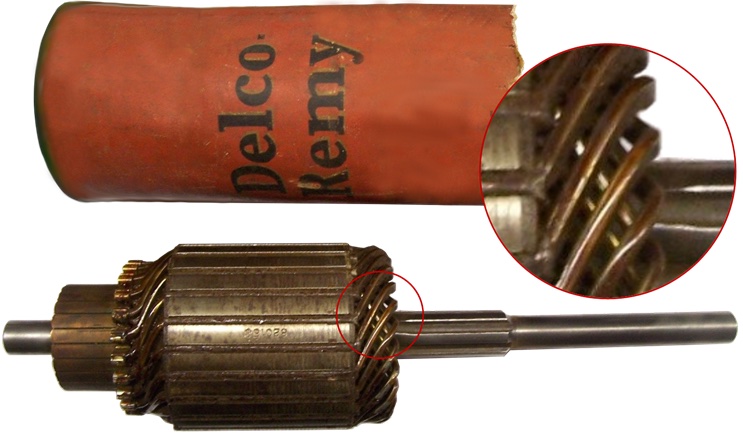}\label{delco_start}}\hskip 0.15cm
\subfloat[]{\includegraphics[height=0.33\columnwidth]{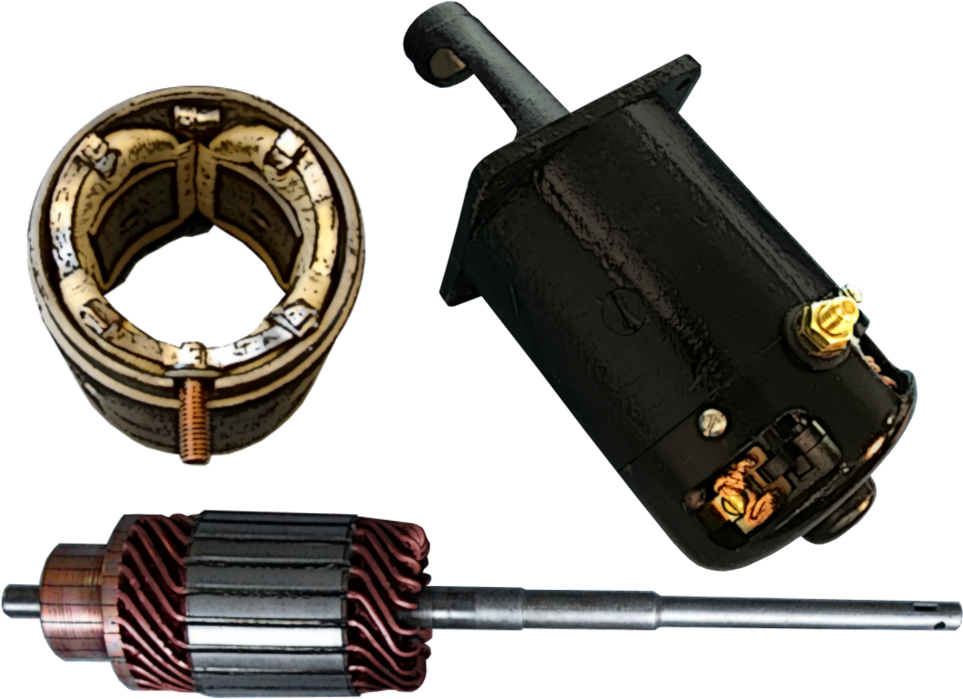}\label{ModelT_complete}} \subfloat[]{\includegraphics[height=0.42\columnwidth]{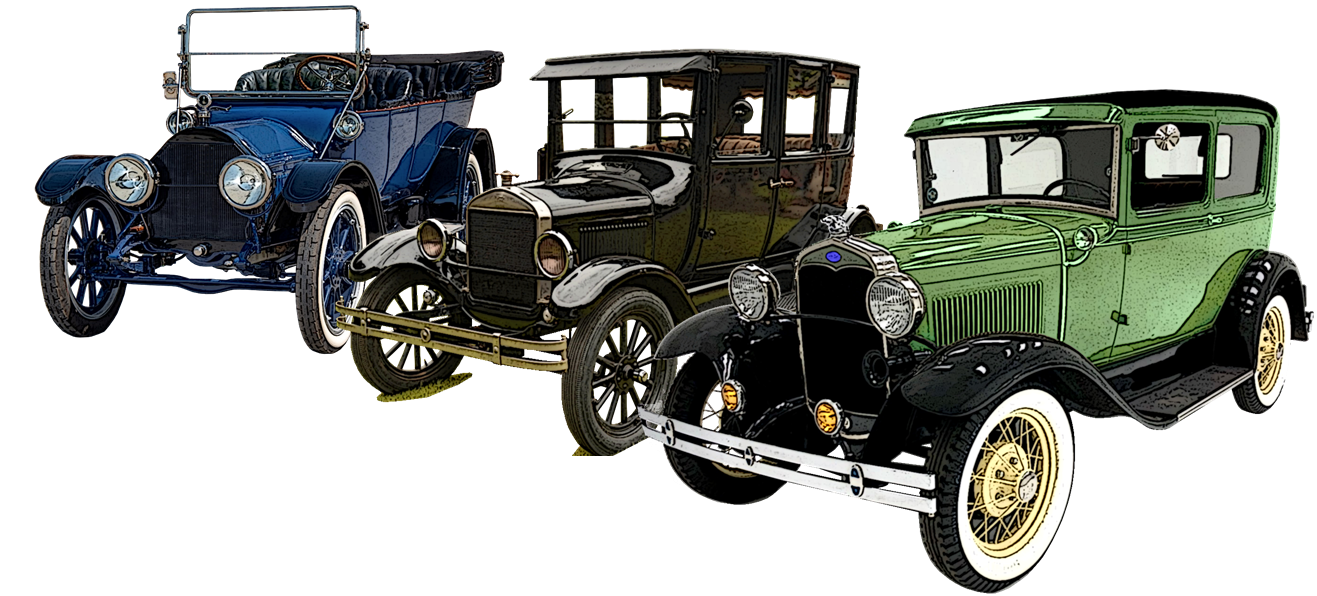}\label{e_started}}\\ \vspace{0.1cm}

%\subfloat[]{\includegraphics[height=0.6\columnwidth]{figures/Cadillac30_Collage_A.png}\label{CadillacCollage}}\\ \vspace{-1cm}
\subfloat[]{\includegraphics[height=0.23\columnwidth]{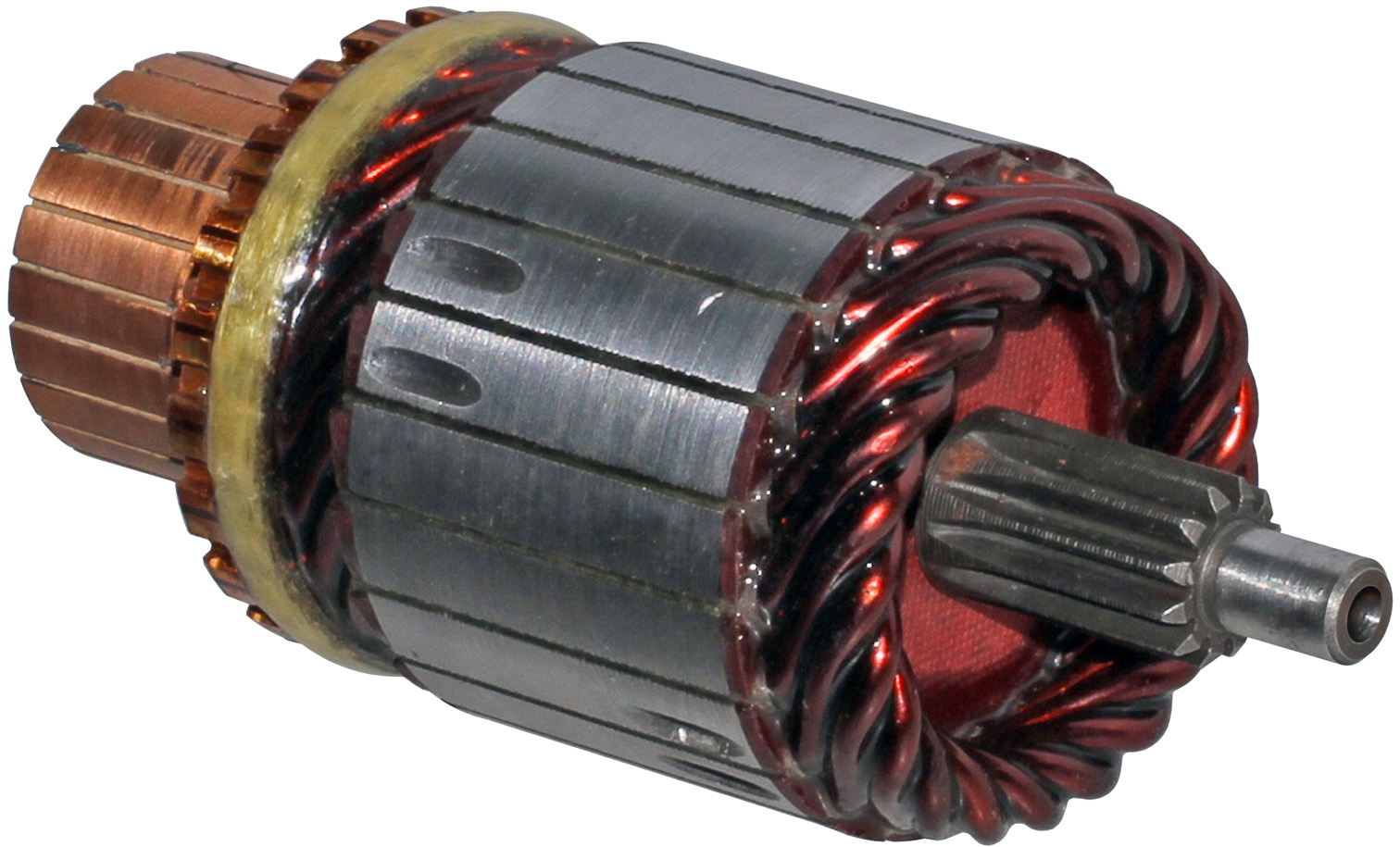}\label{remy}}\hskip 0.3cm
\subfloat[]{\includegraphics[height=0.324\columnwidth]{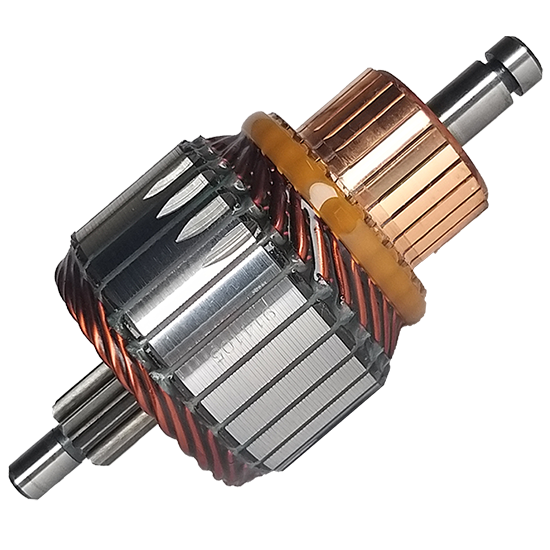}\label{bosch}}%\hskip 1cm
\subfloat[]{\includegraphics[height=0.12\columnwidth,angle = 45]{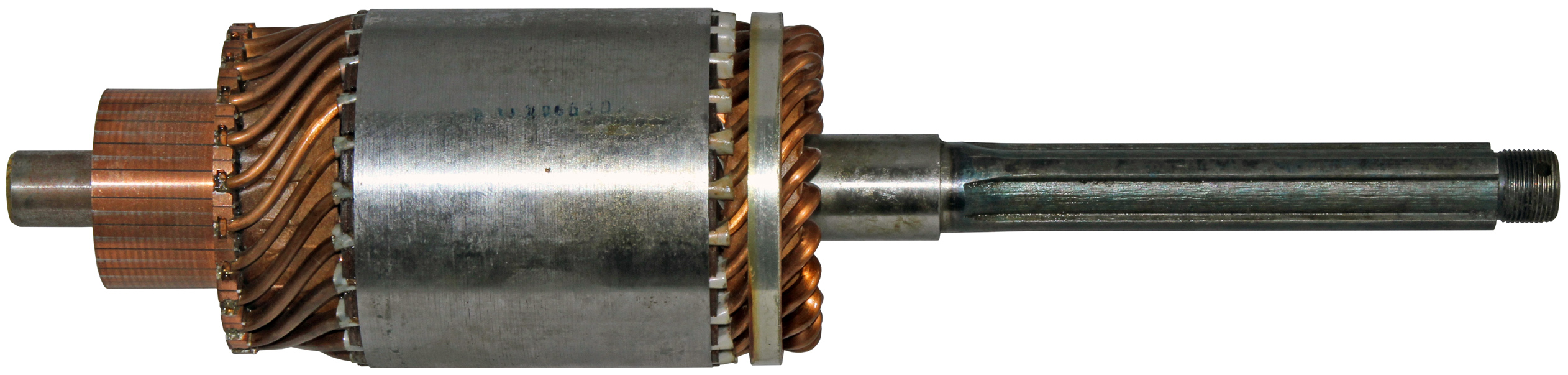}\label{lucas}}%\hskip 1cm
\subfloat[]{\includegraphics[height=0.324\columnwidth]{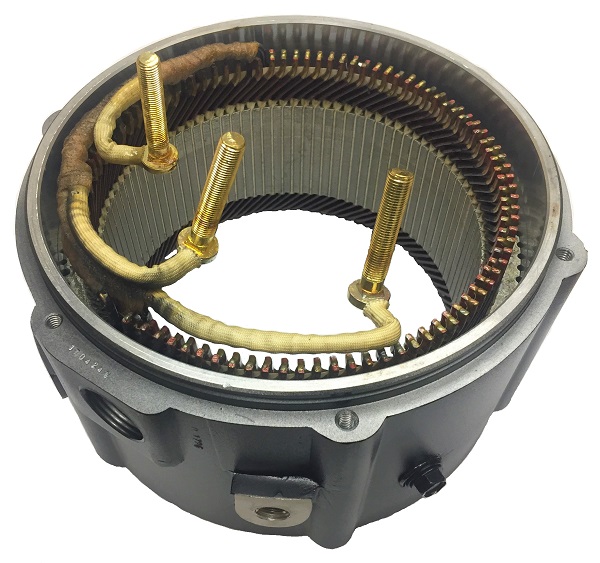}\label{50dn}}\hskip 0.5cm
\subfloat[]{\includegraphics[height=0.37\columnwidth]{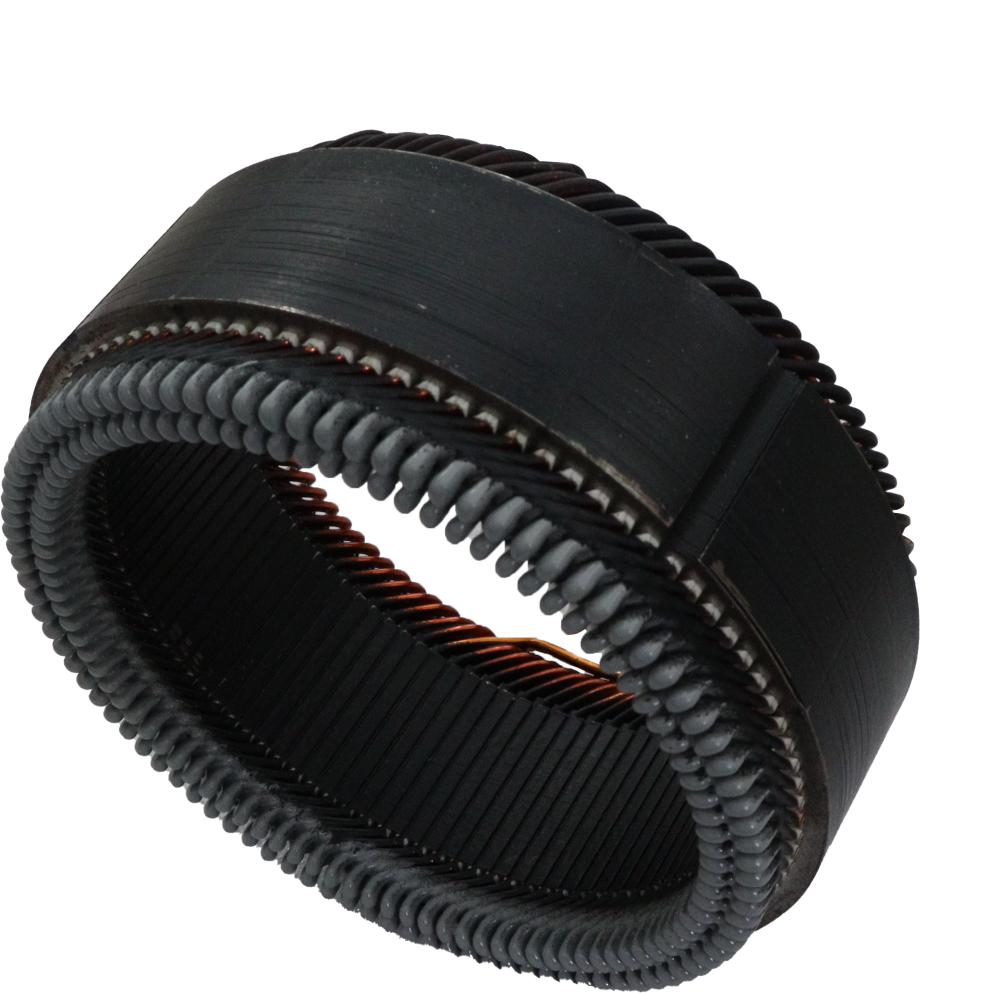}\label{denso}}

\caption{(a) Contemporary advertisement of the Cadillac Model Thirty as one of the first series vehicle with an electrical starter (Reference No.\ 33, right-most picture) for the four-cylinder engine (the engine compartment in C.\ Kettering's patent shown center-left almost perfectly matches the Cadillac's \cite{Kettering:1911:US1150523}); (b)  Bar-wound starter armature for Buick from the early days of electric starters, (c) stator, armature, and casing for a Model T for retro-fitting. The bars have obviously already hairpin shape where the bent side neatly guides the pins around each other using two radial levels, which is so characteristic in modern hairpin motors (see magnification). (d) From left to right: Cadillac Model Thirty; the likely more widely known Ford Model T in the middle, which in its standard configuration was with hand-cranking but allowed already retro-fitting with an electrical starter (see the photo and notice the absent hand crank), and the successor Ford Model A, which had an electrical starter installed in its standard configuration. (e) Delco Remy 1.7 kW 12 V PG260M PMGR Starter Armature (f) 1.1 kW 12 V Bosch Starter Armature. (g) Lucas 1.1 kW 12 V M45G Series Starters Armature.  The development of bar windings in dc armatures and ac motors was closer as nowadays. Early ac machines, as those of GE had the ac winding on the rotor and therefore closely resembled dc armatures but with three-phase slip rings instead of a commutator. Accordingly, the windings could be practically the same. More familiar, however, is nowadays the rotating field on the stator with the static poles on the rotor for synchronous machines as in the (h) Delco Remy 50 DN  270~A/24~V 72-claw-pole alternator, which was first introduced in the 1950s. (i) Recent Denso DAN930 alternator.}\label{starters}
}
\end{figure*}

Hairpin and other bar windings for small drives in the modern sense %as segmented bar windings
experienced a major thrust in the 1910s and more so the 1920s, as a consequence of rational production and growing power densities of the young automobile industry \cite[p.\ 1230, Figs.\ 1519--1522]{Hawkins1917}.
Bar-wound dc armatures were, for example, a good match to provide the high torque density required for electrically started automobiles.
A number of contributions of Charles Kettering and Vincent Apple, both from Ohio, shaped the development of modern automobiles (including electric motors), given their numerous influential inventions \cite{Kettering:1911:US1150523, Kettering:US1231264, apple1, Apple:1912:US1091420, Kettering:US1066432, Kettering:US1171055, Apple:US748334}.  % Beckman/apple1: https://www.daytonhistorybooks.com/manwhorivaled.html
Dayton Engineering Laboratories Company (Delco) along with Remy Electric Company jointly developed technology and manufacturing facilities that set the standard for today's production of starter armatures and triggered a massive incubation phase for novel solutions in the entire industry (see Fig.\ \ref{starters}) \cite{US1544623A,US1555931,US1661344A,US1721810A}.\footnote{Both companies were merged after their acquisition by General Motors in 1918.}  

From today's perspective, starter motors seem to have been the perfect opportunity for bar windings at the right time, when they were struggling with the calamity of ac effects in more conventional motor drives. Starter motors have a relatively low speed and therefore mostly low-frequency armature current content but require massive torque density. Furthermore, manufacturing needs to be highly rationalized to limit cost. Finally, efficiency is secondary due to short usage. As a solution to the high injury risk of hand cranks, electric starters quickly became the industry de-facto (Fig. \ref{starters}). Whereas the Ford Model T was initially without electric starter, Delco soon offered a device for retro-fitting Model Ts (Fig.\ \ref{starters}\subref{ModelT_complete}). General Motor's 1912 Cadillac Model Thirty offered such starters already as a standard configuration based on Kettering's patent \cite{Kettering:1911:US1150523}. In 1927, the release of the Model A, the second biggest success for the Ford Motor Company, included an electric starter as the standard configuration. Figure \ref{starters}\subref{e_started} displays these early examples of electrically started automobiles.

Some of the developments of V.\ Apple have lasting impact and may serve as examples for this rapid technology development. His one- and two-layer winding design for dc armatures with single-turn bars anticipated modern hairpins including the insulation concept using specifically folded paper lining for the slot walls and separating pins (Fig.\ \ref{fig:insulationpaperhist}) \cite{US1224518}; he contributed elegant transposition patterns for more than two bars supporting both lap and wave windings, as well as the combinations of both (such as frog-leg windings) in a single armature \cite{US1826295A}; and he developed the I-pin concept of half hairpins further, for instance, for solder-free dual-commutator armatures \cite{US1789128A}. In the aftermath of these developments, bar-wound technology became a de-facto industry standard for automotive dc starter armatures and is widely used by practically all major equipment suppliers in that business as displayed in Figs.\ \ref{starters}\subref{remy}, \ref{starters}\subref{bosch}, and \ref{starters}\subref{lucas}, with examples from Delco Remy (now BorgWarner), Bosch, and Lucas Industries.
%, eventually becoming the dominating technology for starters from almost all major suppliers.
The conductors of most modern starter armatures have typical hairpin shape with two legs and use round or, less frequently, square, or rectangular strip conductors with often just one, sometimes two conductor per slot.

%agregar citas de: US1544623_1922 US1555931 US1238959 US1661344_1926 US1661332_1924 US2359384_1940 US1742190_1927 \textbf{_US1721810_1924}

Given the cost pressure already building up in the early automotive industry, also manufacturing had to quickly be highly rationalized and professionalized to levels that most engineers from the 21st century would not expect. The assumption that modern manufacturing equipment and processes in the automotive world are the pinnacle of production technology and way ahead of the primitive past does not hold. The tools for hairpin manufacturing from a hundred years ago already introduced most of the techniques as well as tricks and appear hardly distinguishable from modern machines (Figs.\ \ref{fig:bendingmachinesA}). Although old books may still present rather manual tools and formers for prototyping and low-quantity or bespoke motors may look simple (Fig.\ \ref{fig:manufacturingCNC}), the need for these automotive starters at relatively large numbers led to the two fundamental ways to form hairpins through either bending them with full control over the cusp shape in a widely automated tool with a bending mold (Fig.\ \ref{fig:bendingmachinesA}) as well as bending pins through rotating two discs with the cusp following passively (Figs.\ \ref{fig:bendingmachinesA} and \ref{fig:bendingmachinesB}). Both techniques are still the dominant methods to manufacture hairpins to this day.

Automotive ac machines with hairpin stator winding, particularly synchronous machines, on the other hand, did not play a major role until the second half of the 20th century for heavy-duty applications \cite{worley,https://doi.org/10.1002/eej.22522}. An early example is the Delco Remy 50 DN, a 215~A / 12~V claw-pole alternator for buses introduced in the late 1950s, displayed in Fig.\ \ref{starters}\subref{50dn} and sometimes considered to be the first modern series-manufactured ac hairpin machine. Compared to wire-wound alternators, this typically oil-cooled machine had a high slot number of 72 for a two-layer wave winding, which resembles modern traction machines in electric vehicles. Delco Remy reported an efficiency even below 50\% for the original design in the patent literature, due to severe ac losses in the thick conductors at high speed (6500~r/min) \cite{US7633198B2,US7034428B2}. Since then, their design has been improving, partly re-learning and re-discovering how to make hairpins efficient, \emph{e.g.}, short-pitching, keeping the lower part of a slot close to the air gap free (apparently wasting space, unused for bars and current, but still reducing the loss overall) against rotor stray flux with its massive impact on large-cross-section conductors, and forming the crowns of the teeth in partially closed slots.
These improvements allowed further ac loss reduction and increased voltage capabilities, for instance, the updated 50 DN version reaches 78\% in efficiency and is rated for 270~A / 24~V \cite{brochure50dn,brochure50dnB}.

Nowadays, bar-wound, mostly hairpin, alternator stators (\emph{e.g.}, for 12~V) are available from various manufacturers. Among the earlier ones were Denso \cite{US5965965, US6208060} and Valeo \cite{US7126245, EP1554794, US7116020};
% also Denso A6352
the relatively recent Denso DAN930, for instance, has four bars per slot, a triangular shape on the welding side, and relatively round end turns on the other side, the one with the hairpin vertex or cusp as displayed in Fig.\ \ref{starters}\subref{denso}. Most interestingly, the two pairs of bars, each of which alternates between their two positions in the slot on the welding side as in most other hairpin wave windings, are interwoven by bending the outermost bar of each slot over all others to become the innermost on the vertex side; the next slot can alternate between neighboring bars again. Thus, a current is running through all four bar positions in the slots. This transposition scheme might appear novel but actually very closely follows the design of V.\ Apple some hundred years earlier (see Fig.\ \ref{fig:transpositionshapes}) \cite{US1826295A}.
% \emph{e.g.}, Denso

\section{High-voltage Bar-Wound Motors for Traction}
%,Baba:2011:DE12012201698 rather DE102012201698
%Likewise 

\begin{figure}[!t]
    \centering
    \includegraphics[width=\columnwidth]{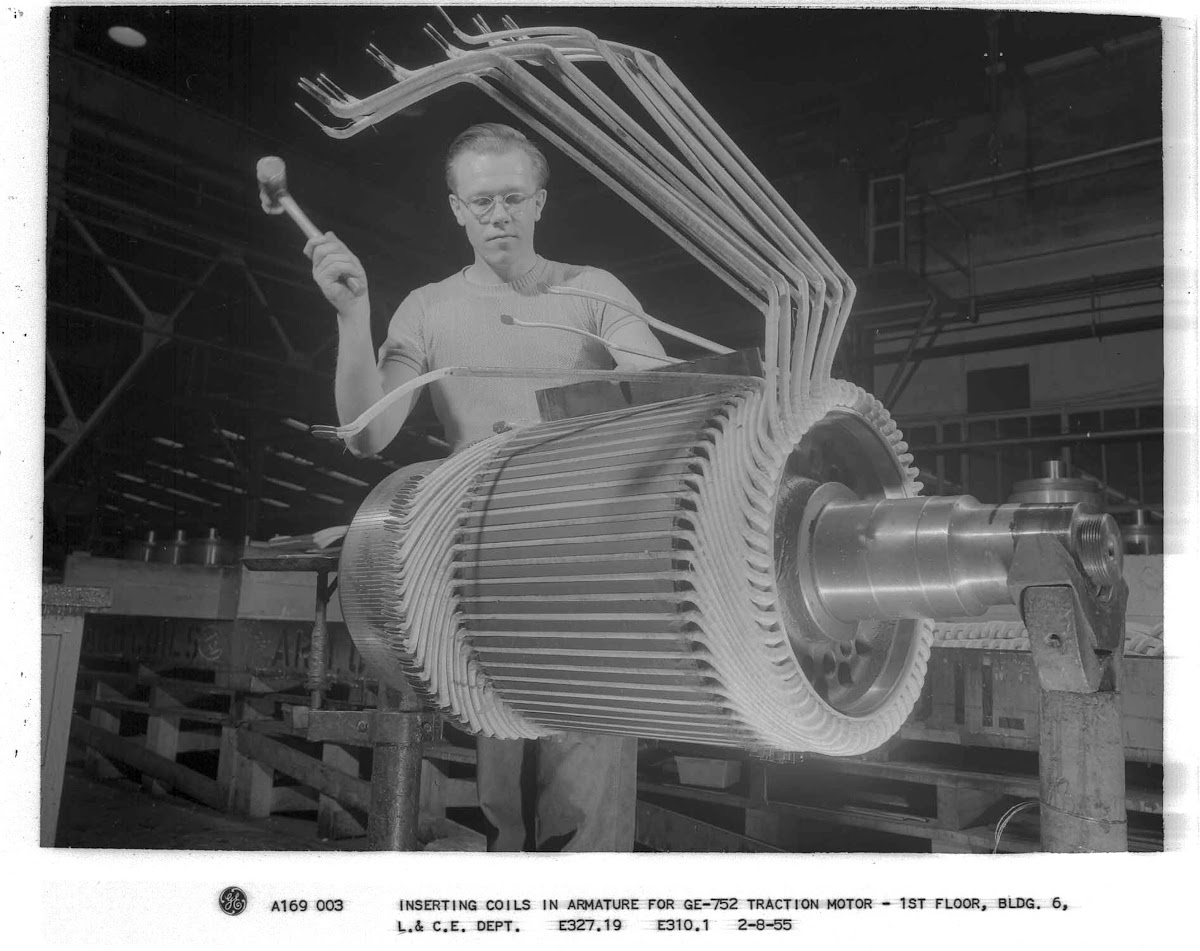}
    \caption{Armature bar insertion for GE~752 showing the characteristic pair-wise transposition at the end-turns known from modern hairpin machines.}
    \label{ge752}
\end{figure}

%hence resembled earlier train and streetcar drives
In the 1940s when the automotive world knew bar-wound machines for alternators and starters mostly, General Electric explored this winding technology for higher-power diesel-electric locomotive traction. The GE~752 motor series displayed in Fig.\ \ref{ge752}, for instance, was a higher-voltage, high-torque, and wide-speed-regulation-range dc traction motor. By 1950, it became GE's basic motor for large locomotives (American Locomotive Company's passenger (PA) and freight families (FA)). Improvements in insulating materials, specially for high-temperature environments led to lighter and more powerful traction systems, since the combination of solid insulators with synthetic varnish led to thinner and improved insulation of the conductors, besides improving the fill factor and the heat-transfer coefficient \cite{5058518}. Consequently, these technological advances also enabled an increase in the voltage range of bar winding machines. Early versions supported 900~A with a rated voltage of 750~V (max.\ 1300~V) \cite{Kirkland:1989}. %\textcolor{red}{\textbf{something is missing between these paragraphs}}
% GE 752 from 1955
Though intensively modified and extended, the GE~752 series exists to this day, mainly used in high-torque drilling applications since better options appeared for traction in the mid 1980s.

Eventually, Delco Remy developed technology for stronger insulation and further increased the voltage ratings for ac machines, enabling a rather new application for automotive traction machines; this development has resulted in distinctive hairpin designs arranged in wave or lap segments---some of which use open slots for insertion---from various manufacturers \cite{US7034428B2, US9130416B2,US9847683B2,US9520753B2,US20200412195A1}. Nowadays, several concepts and jargon terms can be classified under the term bar-wound coils, such as hairpin, single and dual diamond coils, bent and straight concentric coils \cite{Braymer1920,Richter1950}, in addition to coils that were not built with solid conductors, but sub-structured rods (such as the Roebel bar). %\textcolor{red}{\textbf{Plug-in coils or hairpin coils are often used for traction drives since they can easily be handled and produced}}.

\begin{figure*}[t!]
\centering{
\subfloat[]{\includegraphics[height=0.35\columnwidth]{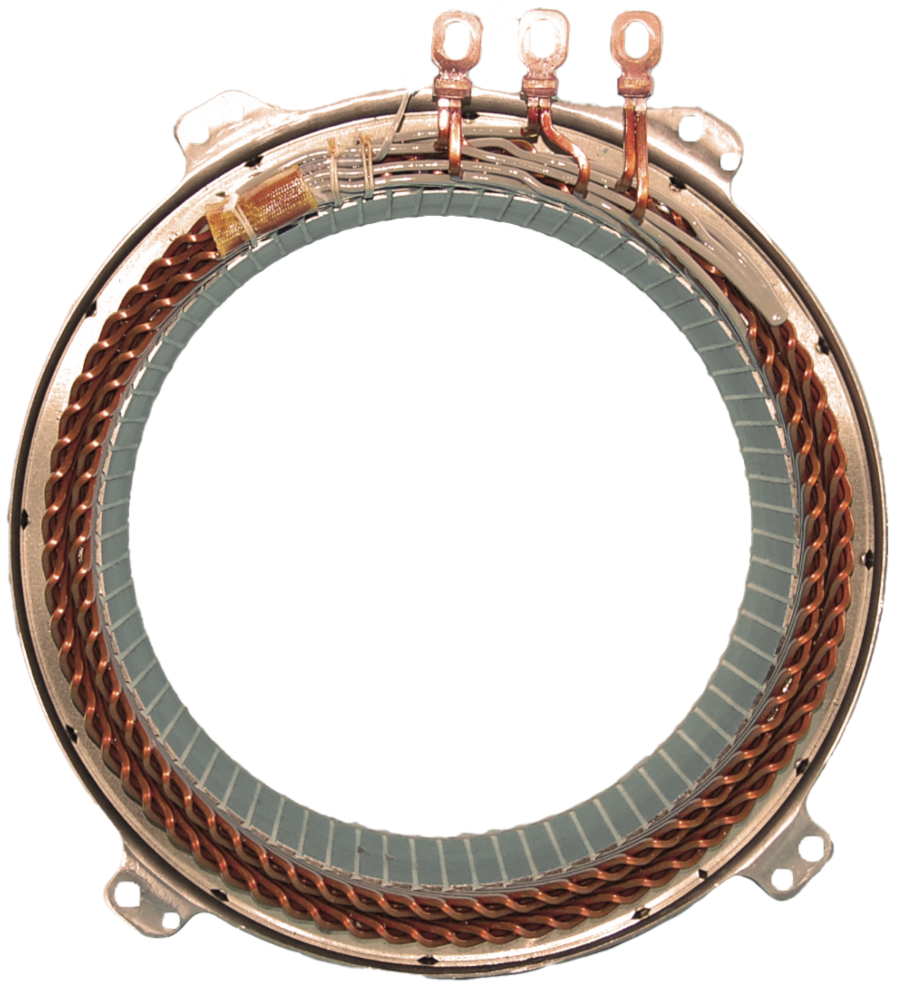}\label{volt}}\hskip 3mm
\subfloat[]{\includegraphics[height=0.35\columnwidth]{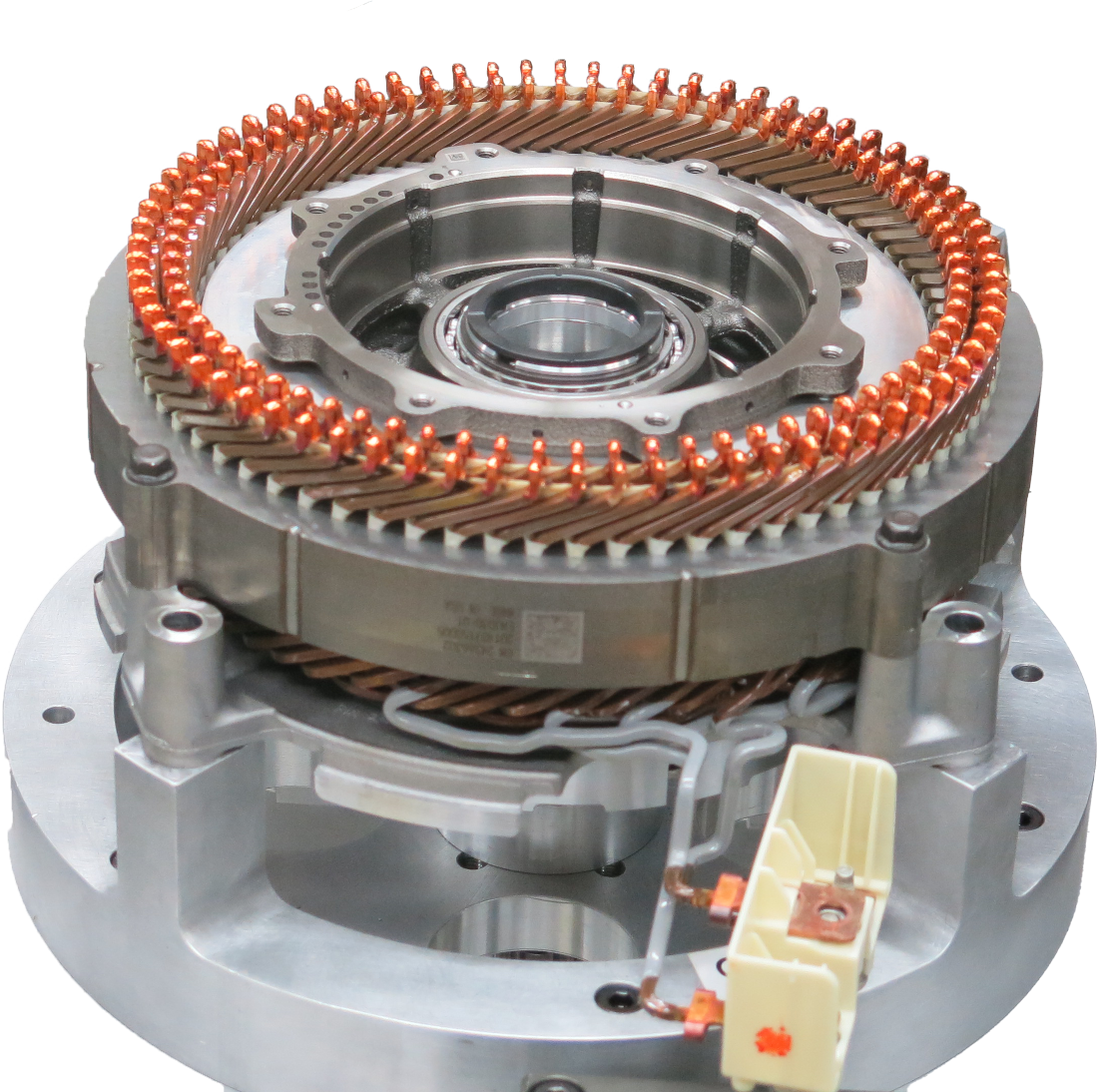}\label{volt2}}\hskip 3mm
\subfloat[]{\includegraphics[height=0.35\columnwidth]{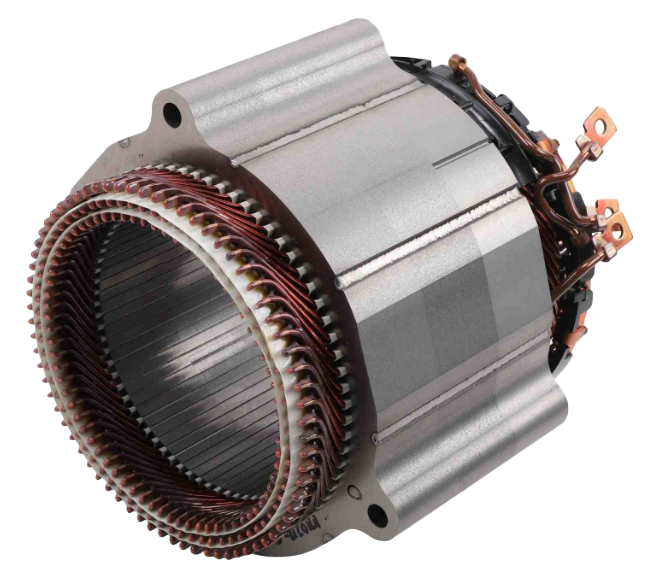}\label{bolt}}\hskip 3mm
\subfloat[]{\includegraphics[height=0.35\columnwidth]{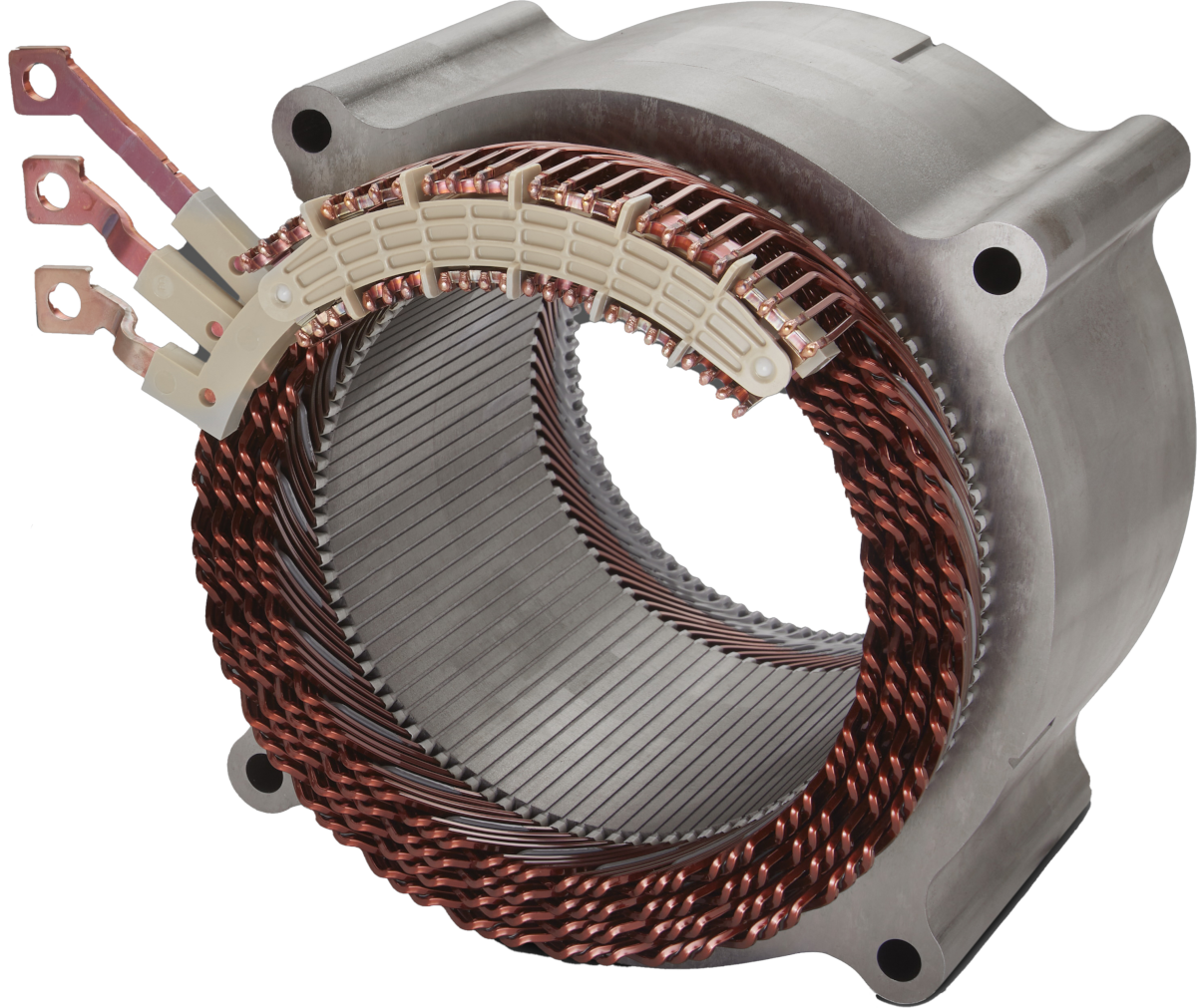}\label{ultium}}\hskip 5mm
\subfloat[]{\includegraphics[height=0.2625\columnwidth]{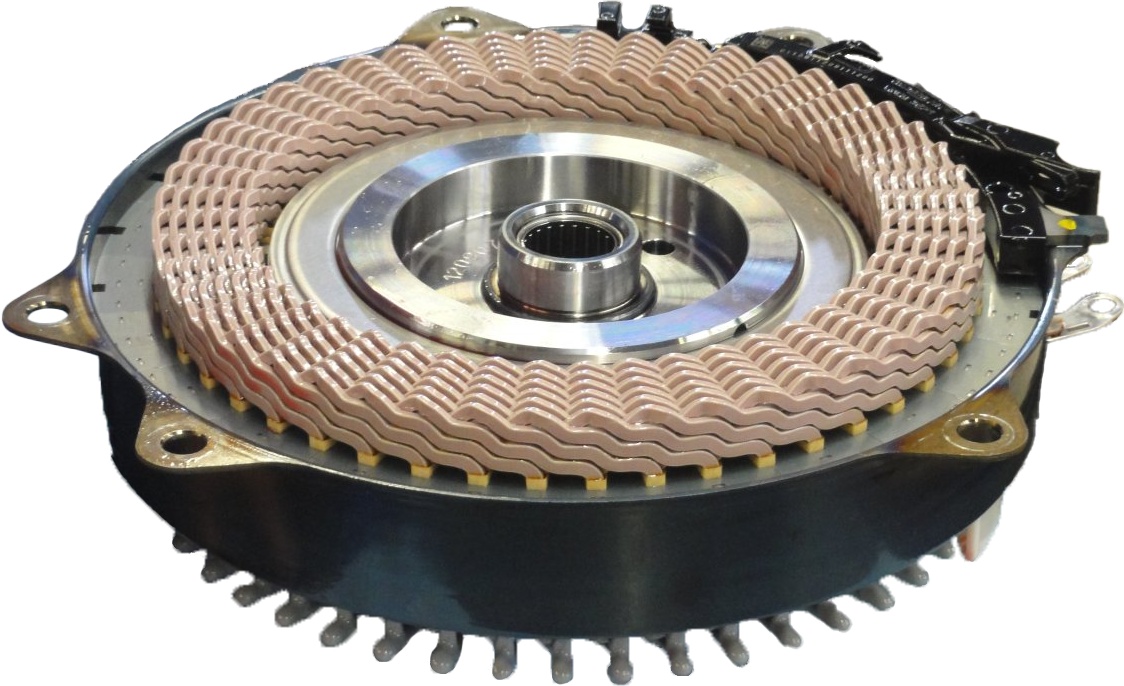}\label{prius}}\hskip 5mm
\subfloat[]{\includegraphics[height=0.35\columnwidth]{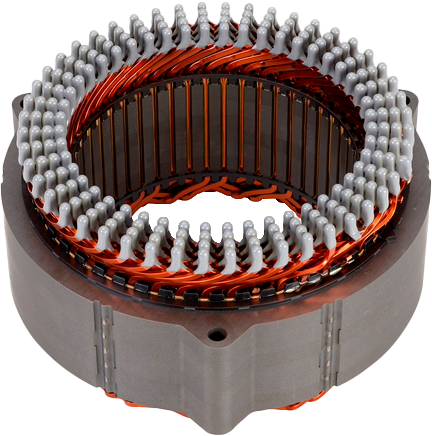}\label{prius2}}\hskip 5mm
\subfloat[]{\includegraphics[height=0.35\columnwidth]{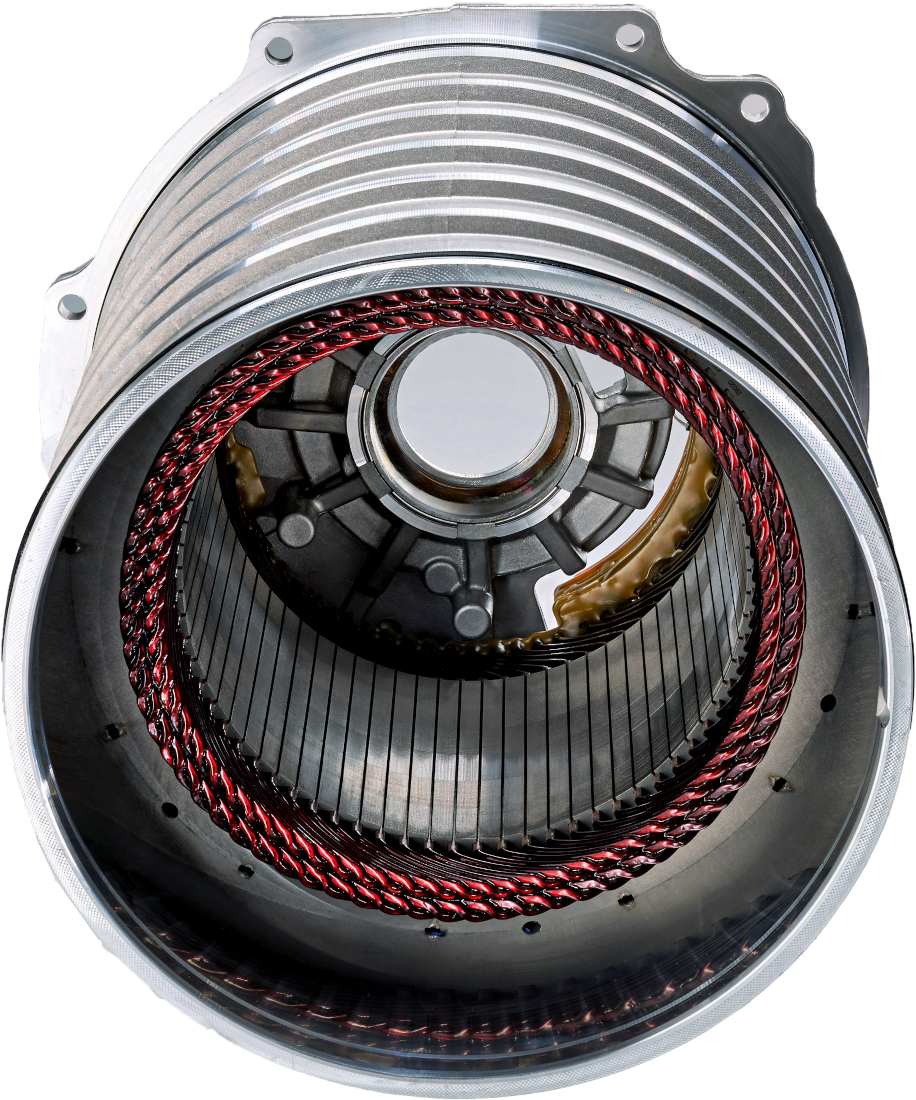}\label{taycan}}
\caption{State-of-the-Art Hairpin/Bar Winding Machines. (a) Chevy Volt Gen 1 stator  GM. (a) Chevy Volt Gen. 2 Motor  GM. (b) Chevy Bolt stator  GM. (d) GM Ultium rear motor stator  GM. (e) Toyota Prius C P510 Transaxle Motor. (f) Toyota P910 Transaxle Stator. (g) Porsche Taycan stator.}\label{modern1}
}
\end{figure*}

Partly based on Remy's footwork, General Motors laid the ground for the mass-production of bar-wound motors for hybrid-electric vehicles in 2006 and subsequently extended it to battery-electric vehicles. Their widely variable modular concept can serve as the basis for both permanent-magnet synchronous and induction machines \cite{8453870,6841736,7855402}. The maturity of this technology reached the mass market in 2010. The stator presented in Fig.\ \ref{modern1}\subref{volt} belongs to the first generation of the Voltec electric drive system, used in the 2011 Chevy Volt, also marketed as Opel and Vauxhall Ampera \cite{RahmanSAEI}. This design was further perfected in Voltec's second generation in 2016, and then for the fully electric 2016 Chevrolet Bolt, for which the eight-pole stator increased the number of conductors per slot from four to six with a winding scheme optimized for low ac effects \cite{2016-01-1228}, as displayed in Figs.\ \ref{modern1}\subref{volt2} and \ref{modern1}\subref{bolt}. With the latest Ultium drive, hairpin technology recently entered the heavy-duty vehicle segment of General Motors, where the high torque density appears a perfect match in packages of 180~kW and 255~kW for front and rear traction. The drive shown in Fig.\ \ref{modern1}\subref{ultium} is also supposed to power the BrightDrop Zevo, a typical commercial delivery vehicle. The proven advantages and fully automated manufacturing processes resulted in the adoption of hairpin by various auto makers.

\begin{figure*}[t!]
\centering{
\subfloat[]{\includegraphics[height=0.35\columnwidth]{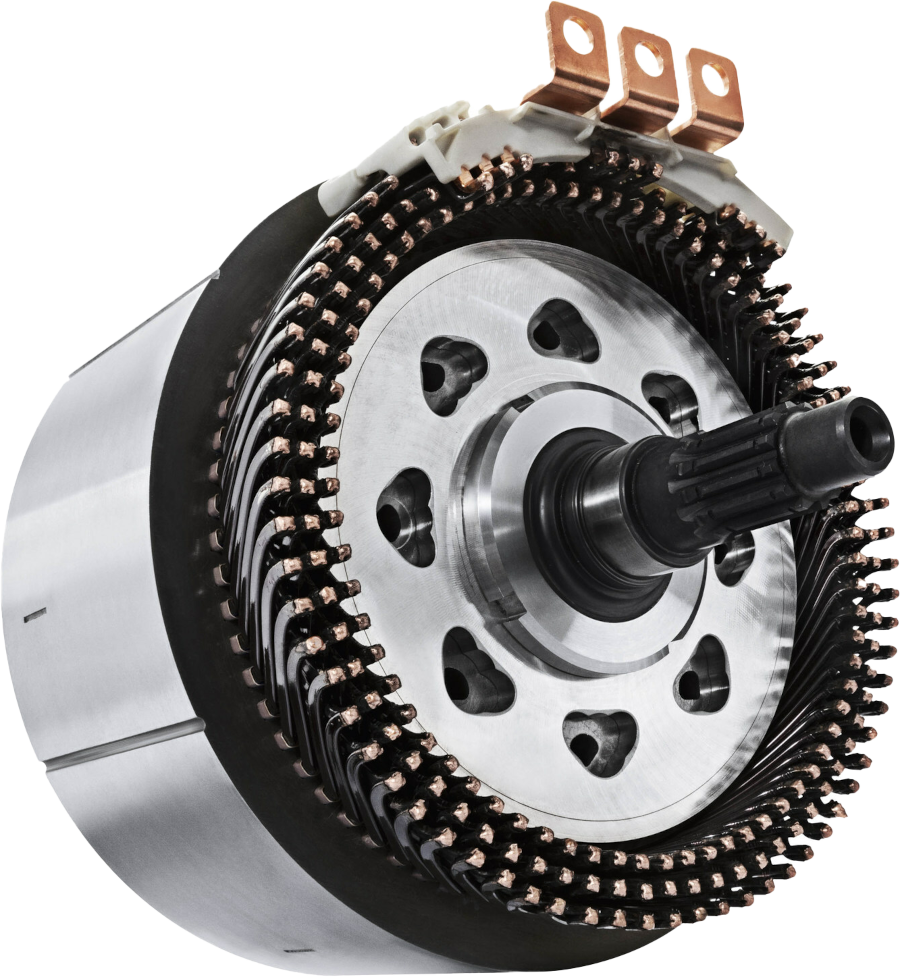}\label{marelli}}\hskip 3mm
%\subfloat[]{\includegraphics[height=0.35\columnwidth]{figures/vw-id7.png}\label{vw3}}\hskip 3mm
\subfloat[]{\includegraphics[height=0.35\columnwidth]{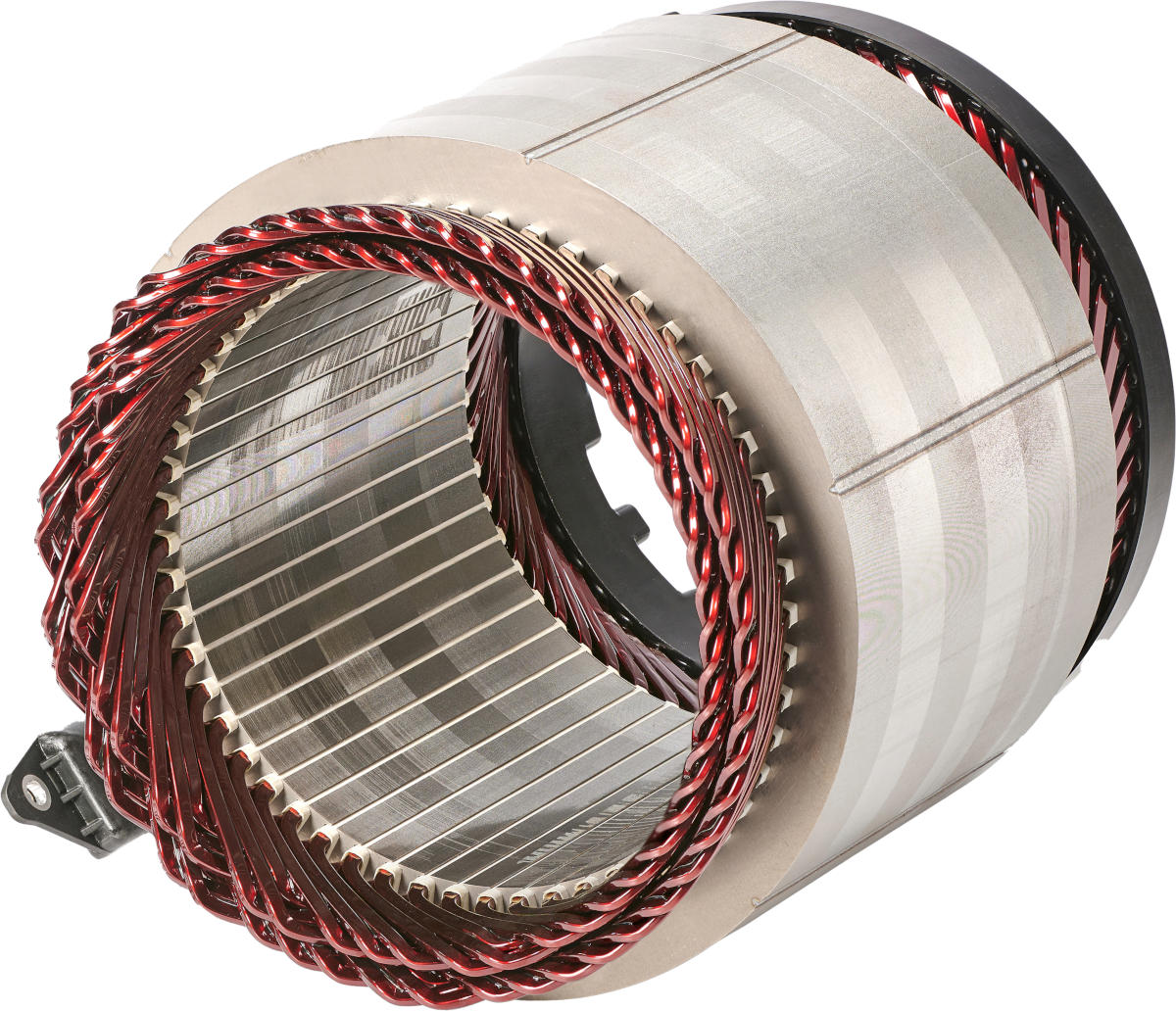}\label{bmw}}\hskip 3mm
\subfloat[]{\includegraphics[height=0.35\columnwidth]{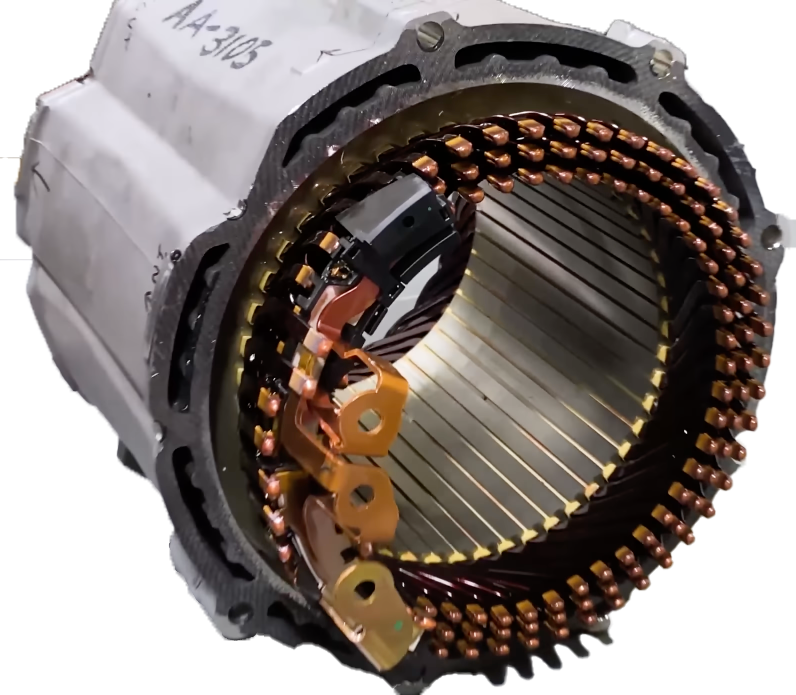}\label{vw1}}\hskip 3mm
\subfloat[]{\includegraphics[height=0.35\columnwidth]{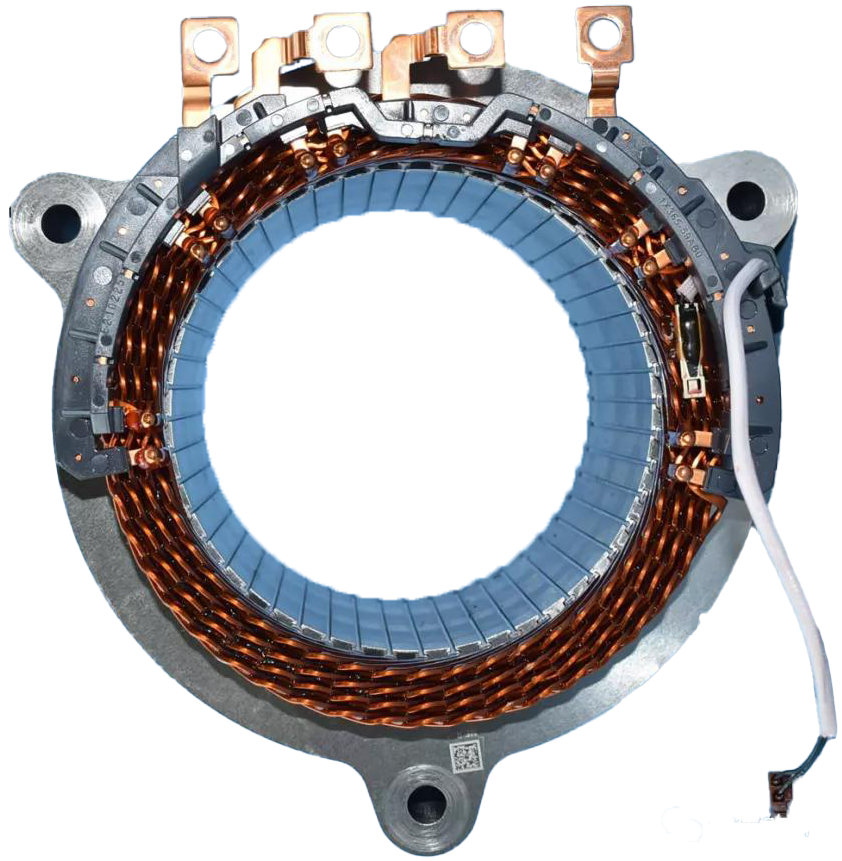}\label{egmp}}\\
\subfloat[]{\includegraphics[height=0.35\columnwidth]{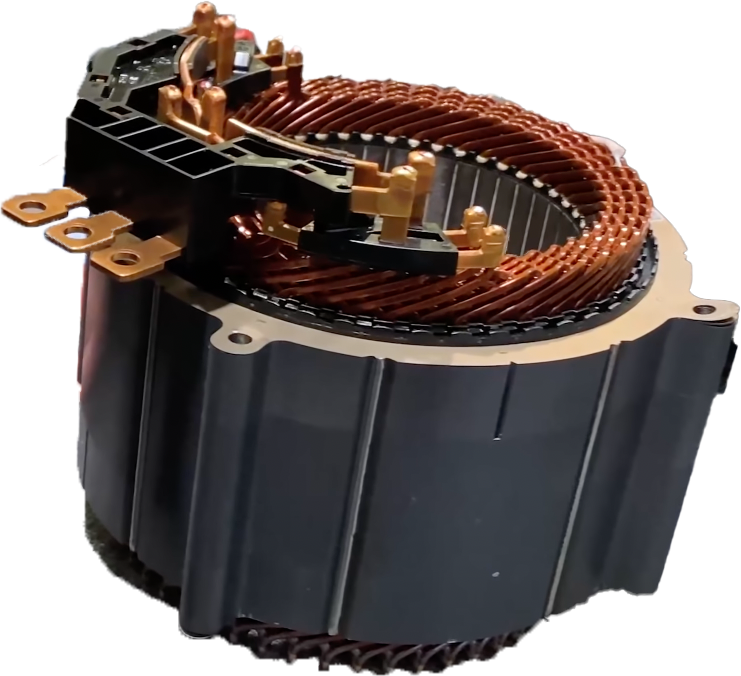}\label{ford}}\hskip 5mm
\subfloat[]{\includegraphics[height=0.35\columnwidth]{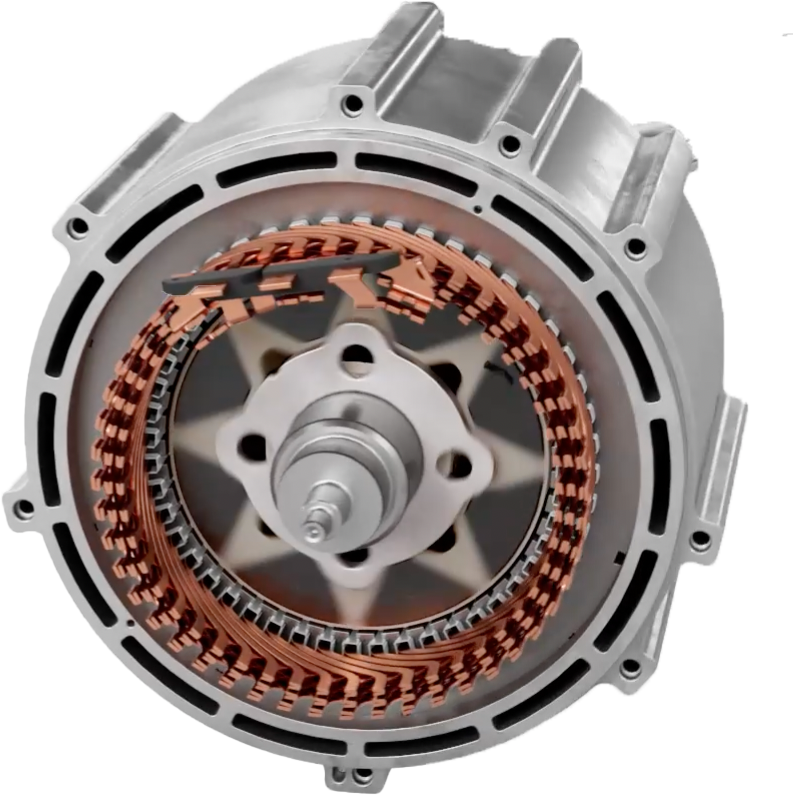}\label{rivian}}\hskip 5mm
\subfloat[]{\includegraphics[height=0.35\columnwidth]{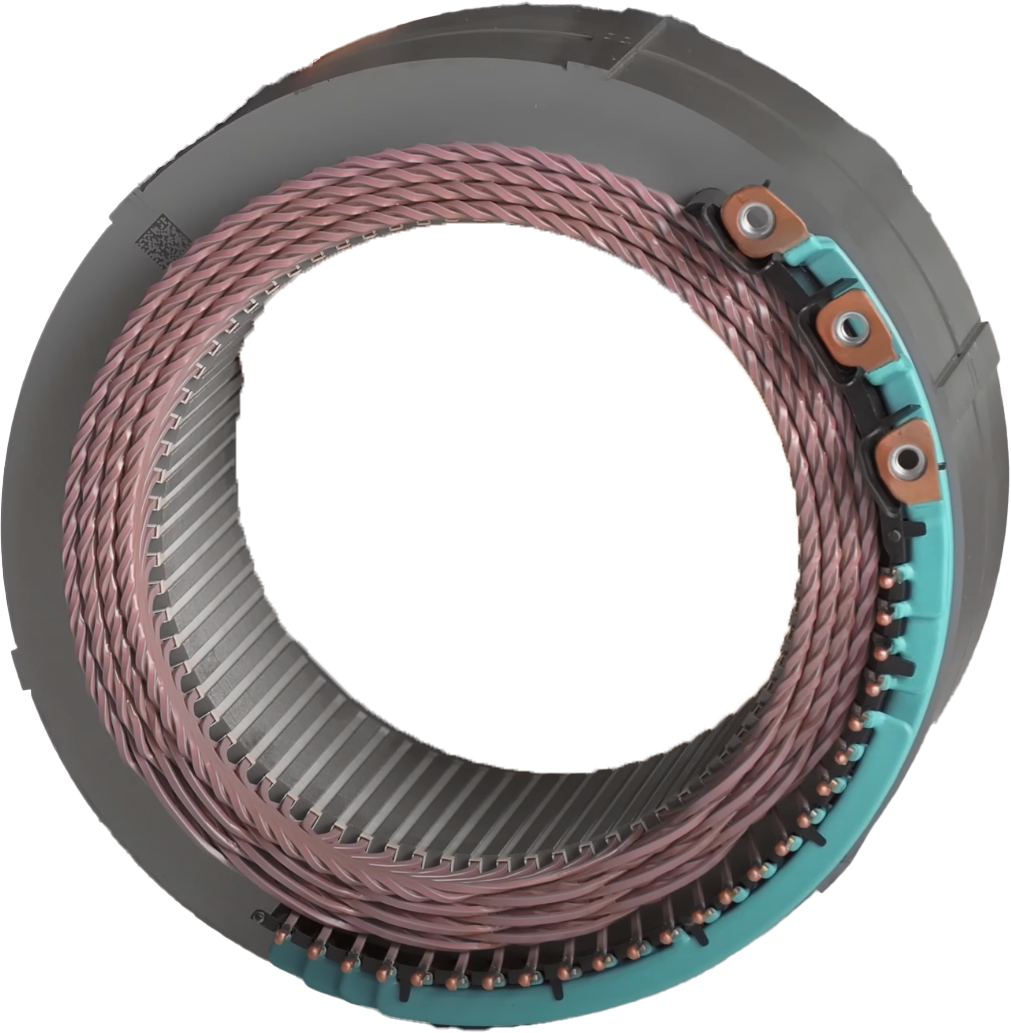}\label{lucid}}
\caption{State-of-the-Art Hairpin/Bar Winding Machines. (a) Maseratti GranTurismo Folgore motor  Marelli. (b) BMW xDrive 5th Gen. stator  BWM.  (c) Volkswagen MEB APP 310 stator  Volkswagen. (d) Hyundai Motor Group Electric-Global Modular Platform (E-GMP) Front Motor Stator  Hyundai. (e) Ford Mustang Mach E stator  BorgWarner
(f) Rivian R1T Motor  Bosch.
(g) Lucid Air Platform Stator  Lucid Motors.}\label{modern}
}
\end{figure*}

\begin{table*}[!tb]
	\begin{center}\caption{Modern Hairpin Motors for HEVs/EVs Part I}\label{table:modern}
		\renewcommand{\arraystretch}{1.45}
			\addtolength{\tabcolsep}{-3pt}
			\begin{tabular}{c}
{\includegraphics[scale=.8]{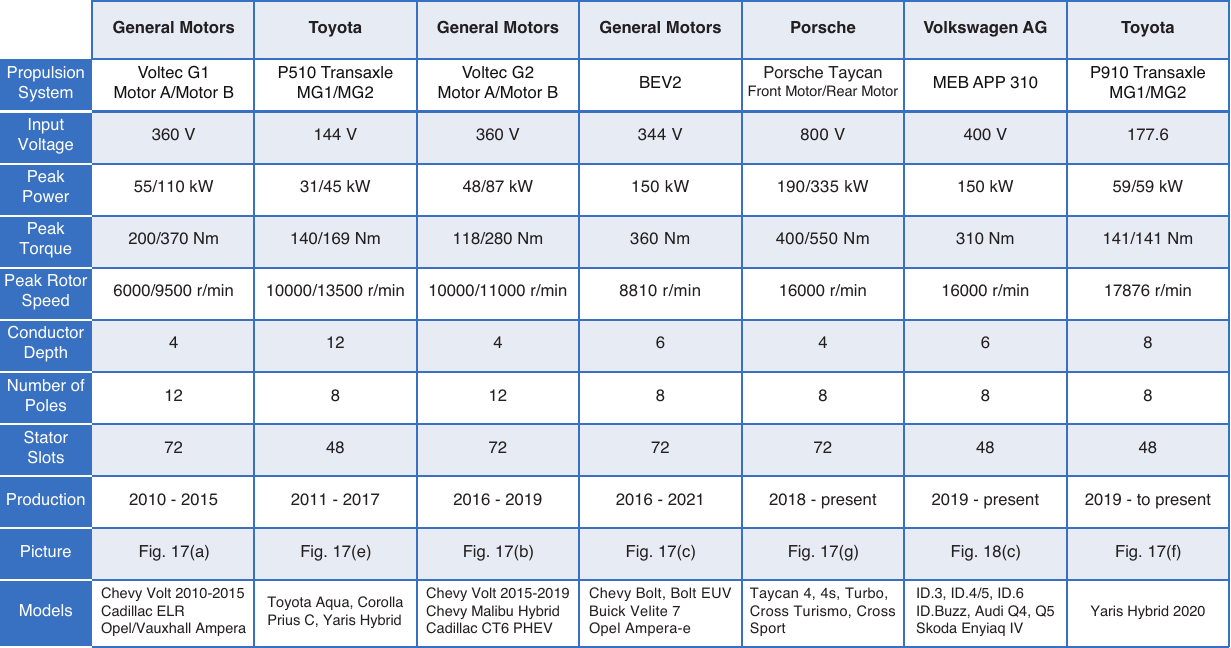}}\\
 						\end{tabular}
 	\end{center}
\end{table*}
\begin{table*}[!tb]
	\begin{center}\caption{Modern Hairpin Motors for HEVs/EVs Part II}\label{table:modern1}
 		\renewcommand{\arraystretch}{1.45}
 			\addtolength{\tabcolsep}{-3pt}
 			\begin{tabular}{c}
{\includegraphics[scale=.8]{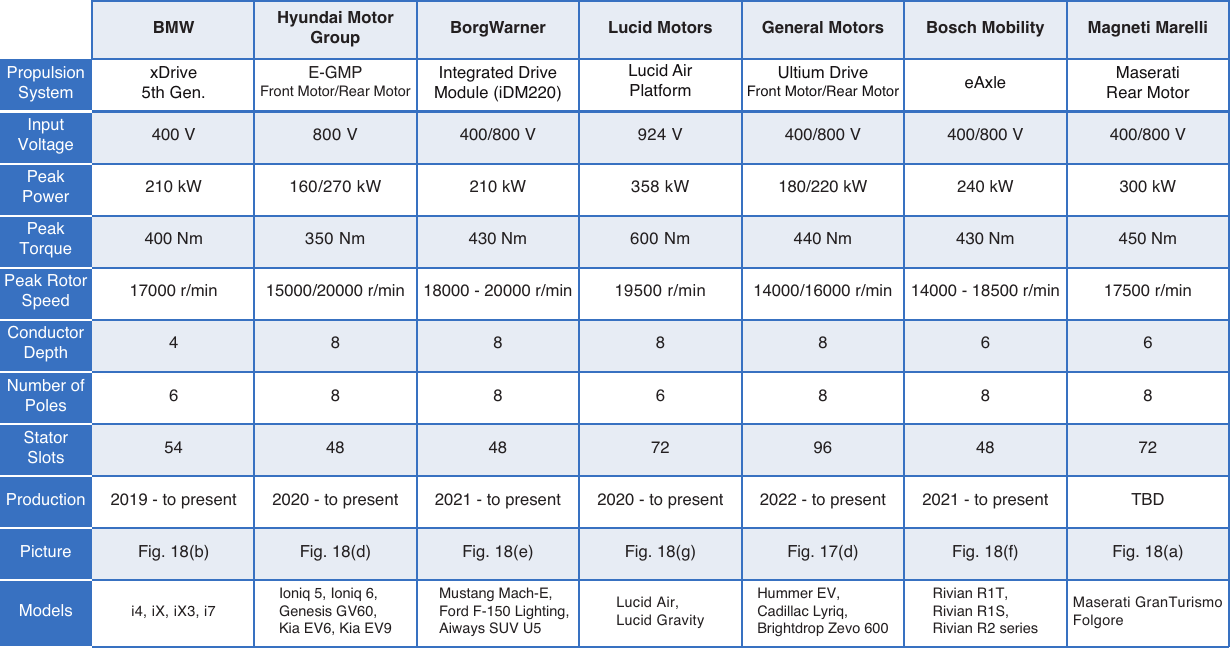}}\\
						\end{tabular}
	\end{center}
\end{table*}
% \begin{table*}[!tb]
% 	\begin{center}\caption{Modern Hairpin Motors for HEV/EV}\label{table:modern2}
% 		\renewcommand{\arraystretch}{1.45}
% 			\addtolength{\tabcolsep}{-3pt}
% 			\textsf{\begin{tabular}{c}
% \multicolumn{6}{c}{\includegraphics[scale=.8]{figures/modern_hairpin_3.pdf}}
% %Intentar corregir el ancho para lograr el mismo efecto de los dc saliendose al otro recuadro
% 						\end{tabular}}
% 	\end{center}
% \end{table*}

Toyota, for instance, features bar windings, though preferably with longer segments than hairpins, in their more recent vehicles, such as the Aqua (Prius) or Yaris, and has turned bar windings into their baseline technology (Figs.\ \ref{modern1}\subref{prius} and \ref{modern1}\subref{prius2}). From Toyota and its in-house suppliers, bar winding technology also diffused to other brands, such as Subaru, so that Toyota forms a seed and breeding ground in Asia. In Europe, on the other hand, Magneti Marelli and Tecnomatic served as pioneers and multipliers. In the meantime, Porsche uses this technology practically exclusively in their battery-electric vehicles in both series as well as motor sports and achieves high torque densities as well as still relatively high speeds in the motors (Fig.\ \ref{modern1}\subref{taycan}). The high torque density may also have been the major motivator behind Ferrari's choice of hairpin motors as well as Jaguar's, \emph{e.g.}, in the I-Pace \cite{Fuchss2019}, and Maserati's in its GranTurismo Folgore (Fig.\ \ref{modern}\subref{marelli}). BMW and Volkswagen might rather have been attracted by the high automation level, which led to the units presented in Figures \ref{modern}\subref{bmw} and \ref{modern}\subref{vw1}. Hyundai on the other hand developed a universal platform denominated Electric Global Modular Platform (E-GMP). The platform displayed in Fig.\ \ref{modern}\subref{egmp} should form the basis for future Hyundai and Kia electric cars. The first vehicle equipped with E-GMP is the Ioniq~5. In principle, the E-GMP is designed as a rear-wheel-drive platform but can also be equipped with a second electric motor on the front axle. After intensive research and development, also Ford implemented hairpin windings in their motors, for example in the Mustang Mach-E or the most recently released F-150 Lightning, shown in Fig.\ \ref{modern}\subref{ford}. % see Kelly's teardown of the Mustang Mach-E GT at 12:53 (https://www.youtube.com/watch?v=y1HulO-EZkc) or here (https://www.macheforum.com/site/threads/geek-out-on-the-engineering-behind-the-mach-e.148/)
% F-150 lightning: Munro https://www.youtube.com/watch?v=1W04m4PioIY
Honda has collected some experience with hairpin windings in their electrified vehicles, particularly the recent generations of the Accord Hybrid. % Honda Accord Hybrid since at least 2017 to my knowledge (smg), see also Kelly's tear-down at 6:20 (https://www.youtube.com/watch?v=QLUIExAnNcE)
%https://cdn.motor1.com/images/custom/munro-associates-12-electric-motors-teardown.jpg
% Clearly Jaguar I-Pace
Some automotive start-up companies have embraced bar-winding technology from the beginning on, such as Rivian in their R1T or Lucid Motors in their Air model, whose rotors are shown in Figs.\ \ref{modern}\subref{rivian} and \ref{modern}\subref{lucid} respectively. A deeper comparison of the different hairpin drive units in the market is provided in Tables \ref{table:modern} and \ref{table:modern1}.

\begin{figure}[!t]
    \centering
    \includegraphics[width=\columnwidth]{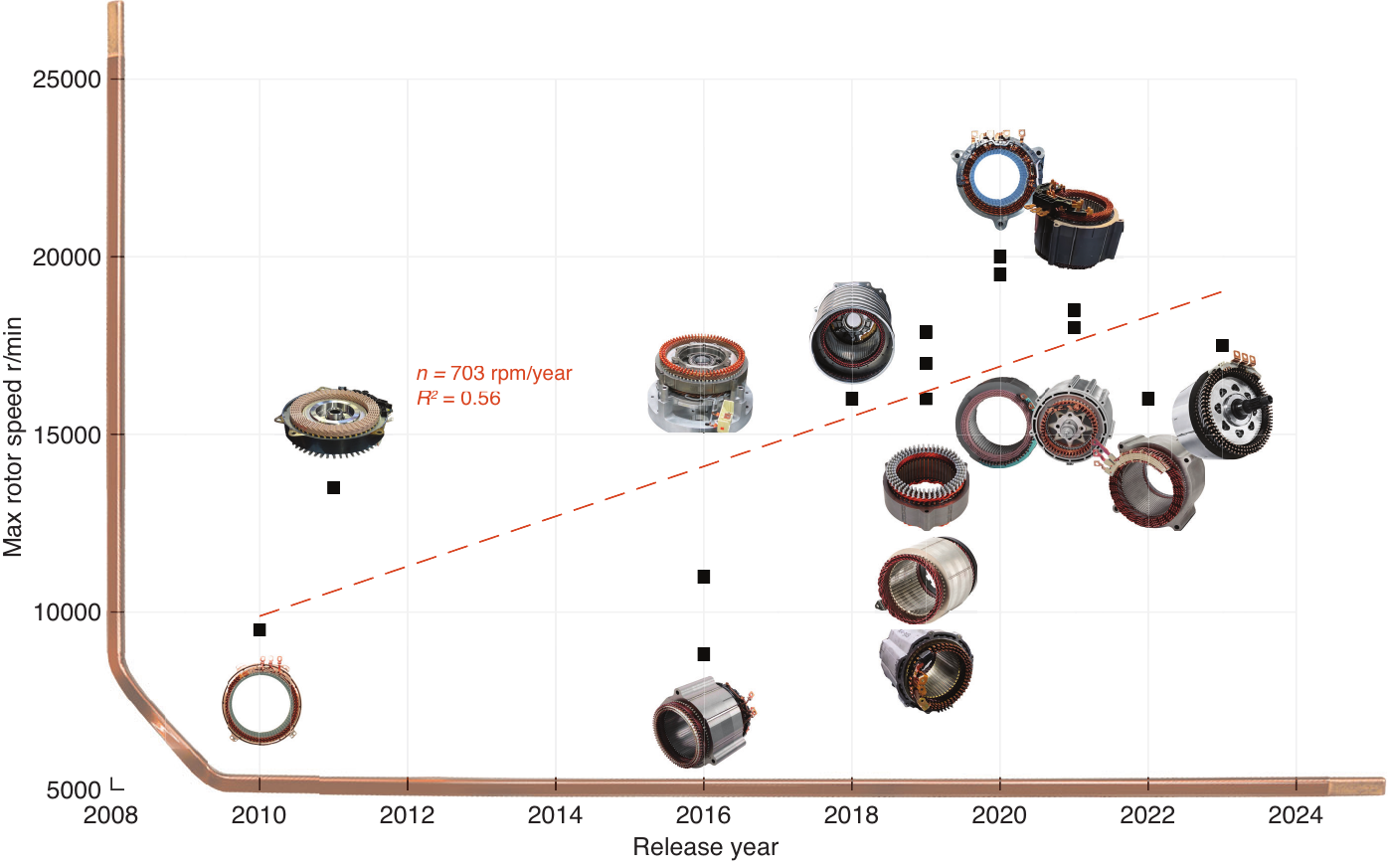}
    \caption{Development of rotational speed of hairpin motors from the first generation of the Chevrolet Volt and Opel/Vauxhall Ampera onwards. The increasing confidence and experience in controlling the high-frequency losses enabled a growth in speed, which is known from conventional motors to enable increased power densities.}
    \label{fig:trend}
\end{figure}

\begin{comment}
rencent things
\begin{itemize}
    \item manufacturing
    \item fully automated bending, placing, twisting, welding vs.\ manual wiring
    \item improving connection from: (1) brazed or clamped (2) welding (pressure/mesh welding, spot welding, TIG welding, fully automated fast laser welding (single color, two-color welding)
    \item manage ac losses through winding techniques: understanding of aspects like short-pitching
    
\end{itemize}
\end{comment}

\section{Latest Technological Developments}
\subsection{Thermal Design and Power Density}
As detailed above, the current trend of hairpin and related bar windings in automotive machines is not a singularity, but can refer to several major leaps in the last 130 years. As a large share of these previous developments may have sunk into oblivion, the ongoing development is consolidating the state of the art, including recovery or even re-invention of older technology. Beyond that, however, the current rush in the automotive industry is also intensively stimulating new developments.

Bar-wound machines have already by design a larger as well as clearer defined interface and thus tighter thermal coupling of the copper to the stator. Whereas most historic bar-wound machines are relatively simple from a thermal perspective and typically use basic air cooling, current research and development is indeed exploring uncharted territory and promises that cooling can be improved substantially. The regular and well-controlled shape of the bars and the boundaries allows reducing the number of thermal interfaces and increasing the contact area. In that context, the simplification of slot liners and their replacement with surface coating, such as powder coating or polymer varnish, promises thinner insulation layers and better contact without air pockets \cite{WO2020104425}. Furthermore, insulation materials on the conductors with higher temperature range than the currently used class-H materials, such as improved polyamide--imides (PAI), polyetherimides (PEI), polyether ether ketone (PEEK), polytetrafluorethylene (PTFE), or imidazole derivatives including polybenzimidazole (PBI), may allow a higher copper temperature for an increased temperature gradient and consequently heat flow \cite{IEC60085,HVinsulationplasticsCarbon}.

The largest gains in power density are expected from more direct cooling approaches of the winding. Instead of previous cooling jackets on the stator back, where the heat had to pass through the thermally poorly conductive stator and multiple thermal interfaces, cooling with liquids as close as possible to the stator winding. For many synchronous machines, the stator winding is the thermally most stressed element at high torque load until at higher speeds rotor losses may become dominant and exceed them. The regular structure of the end turns and the alternating patterns of always two hairpins in a pair as described above provide a large surface for heat transfer if perfused with a liquid medium. The pins on the welding side are anyway spread out for better accessibility and against shorts during welding, and the bending side can be treated likewise to give the cooling medium way and generate a larger, almost fractal contact interface. Importantly, some media may degrade the insulation coating of the conductors so that material-specific compatibility tests are necessary for direct liquid cooling of the hairpin end-turns \cite{osti_10113619}.

\begin{figure*}[!t]
    \centering
    \includegraphics[width=\textwidth]{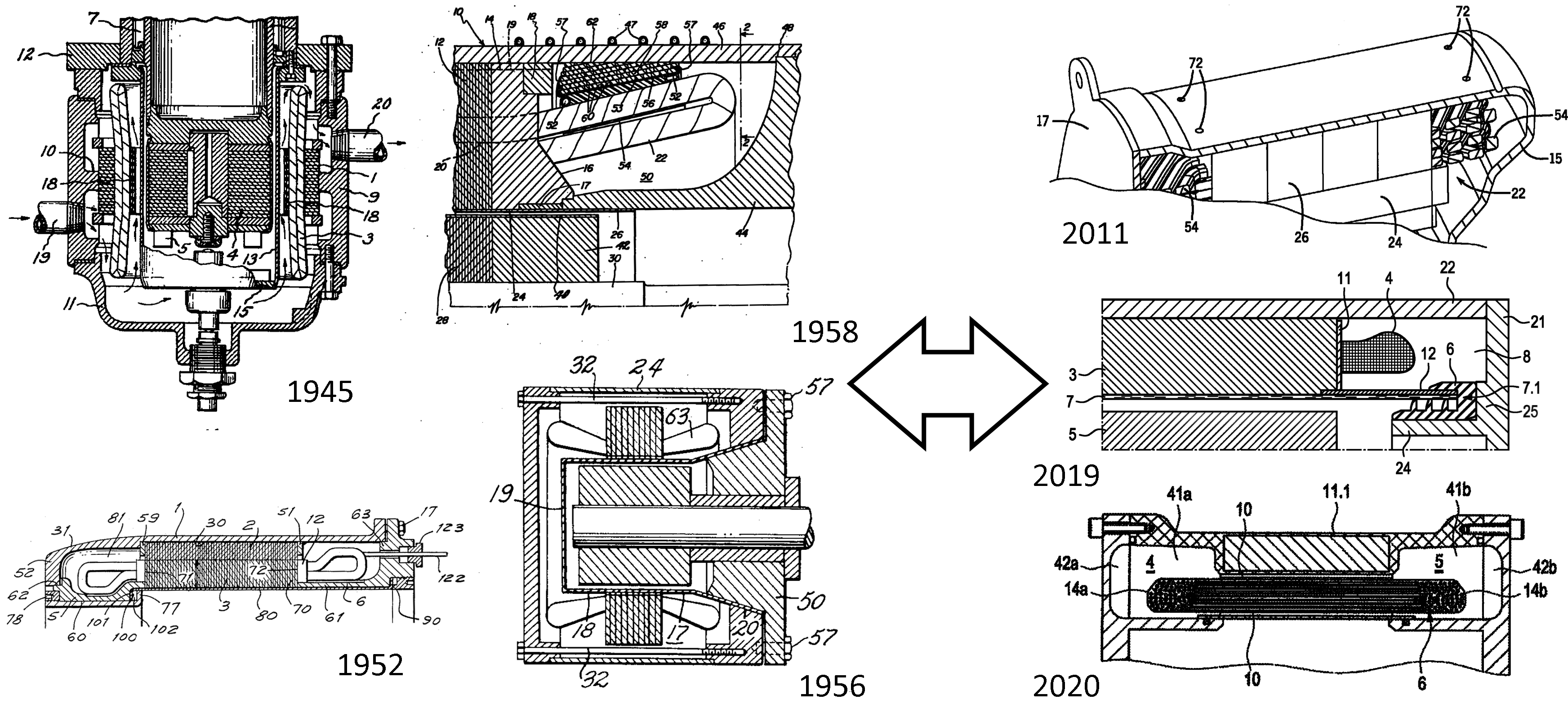}
    \caption{Bar and hairpin windings offer a well defined and sorted end-turns with a large surface area of the conductors for direct liquid cooling \cite{DE102019117373,US8975792,US20200295614}. The necessary separation of a volume around the end-turns to avoid liquid entering the airgap and causing massive friction loss uses techniques well known from canned motors \cite{US2497650, US2727164, US3075107, US2958292}.}
    \label{fig:DirectCoolingHist}
\end{figure*}

Oil as a cooling medium for direct contact may be most obvious as it has been used in many hybrid-electric vehicles where the motor is often integrated into the gearbox and shares the oil circulation \cite{ha2021cooling, US8975792, 10.1115/1.4051934}. Oil is electrically nonconductive, but has the limitations of high viscosity at low temperature as well as large temperature dependence, degrading effects on many polymers used for insulation, and the need to maintain a separate oil circulation system even in battery-electric vehicles without any gasoline engine.

Fluorinert media and alternatives with lower global warming potential such as hydrofluorethers and fluoroketones might be nonconductive alternatives with lower viscosity, but do not necessarily increase the low thermal capacitance of oil \cite{7383287,US10910916}. Whereas such nonpolar and insulating media can serve as a normal coolant in their liquid phase, the use of typical refrigerants, which could absorb large amounts of heat with their vaporization enthalpy, are currently revisited for bar-wound traction machines but need constant minimum pressure and tend to form gas pockets \cite{9899655,9819945}.

A notably higher cooling potential without the need to use a two-phase system with its challenges might be water-based direct cooling of the end turns. Water appears attractive for bar windings due to its high thermal capacitance and manageable viscosity.
%The major challenge is the , although insulation aspects have to be s .
As such water or water/glycol is electrically conductive and can further be corrosive, the conductors as well as the steel lamination have to be fully insulated, \emph{e.g.}, through thin coating or even injection-molded structures \cite{US20200161947}.
%If the con fully insulated, \emph{e.g.}, through coating or even injection molded structure, in principle even conductive media including water or water/glycol with substantially higher thermal capacitance could be used \cite{US20200161947}.
Despite the large potential, however, more intensive research may be required for reliable use.

%{\color{red}
%oil-cooling with open air gap \cite{US8169110}
%}

%Whereas in motors for hybrid-electric vehicles, cooling fluids typically get in touch with the rotor, direct cooling of high-speed electric motors for battery-electric cars needs smarter solutions. 

In hybrid-electric vehicles, the motors of previous car generations were often already integrated in the gearbox, and cooling fluids typically got in touch with the rotor and into the air gap \cite{US8169110}. The speeds of such motors is usually low (engine speeds) and the airgap relatively large. Direct cooling of high-speed electric motors for battery-electric cars, however, needs smarter solutions: The friction loss can reach kilowatts if viscous liquid enters the thin air gap \cite{8507058}. Accordingly, such direct-cooling solutions typically form a separate volume around the end turns to partition it off from the air gap or even the entire rotor (Fig.\ \ref{fig:DirectCoolingHist}) \cite{US8975792, US10938265, DE102016101705, DE102019117373, US20210006132}.
%In principle, such a design reminds of canned motors as used for wet-running pumps, which have been known for decades \cite{doi:10.1243-CannedPump,}.
In principle, most such designs build upon the long experience with canned motors as used for wet-running pumps, which form a thin shell on the surface of the stator towards the air gap and have been known for decades \cite{doi:10.1243-CannedPump,4412212,US3143676}. To avoid large eddy-current loss, most approaches replace the stainless steel can enclosing the rotor and the air gap of conventional canned motors with fiber glass or ceramic fiber compounds. Carbon fiber with its relatively high electrical conductivity and tendency to draw eddy currents can be counterproductive though.

\begin{comment}
\begin{figure}
    \centering
    \includegraphics{}
    \caption{Fig 1 from US 8,975,792? or DE~10~2019~117~373 or US 10,938,265 versus older US~3,143,676 or US 3,075,107 or US 2,958,292 or US 3,192,861. \textbf{\textcolor{red}{Do we still need this figure??}}}
    \label{fig:my_label}
\end{figure}
\end{comment}

%oil \cite{8507058}
%friction loss to kilowatts \cite{8507058}
%Air gap separated \cite{US8975792}
%direct cooling, see US 8,975,792

%Spray vs. channels vs. passive perfusion
%Spray cooling \cite{8848870}
In addition, various ways to improve the contact of the coolant with the windings are studied. Instead of just passive perfusion around a larger open end-turn volume, active spraying of coolant and channels are under development \cite{8848870}.
In-slot cooling, \emph{e.g.}, through cooling channels between conductors \cite{9271026,8785115} as known from large drives \cite{US9246373} or with hollow bars \cite{8648388,9270717,8607655} may be an alternative, but the loss of slot fill, the challenge of cost-efficient, leakage-free injection of the coolant at the end turns, and often high necessary fluid pressure may need more intensive research for competitive solutions. Filling the space close to the airgap with coolant, where rotor stray flux would prevents proper current flow in bars anyway, has been proposed regularly, but to date no reliable routine implementation seems to be in reach \cite{LiuInSlotCooling,9271026}.

%replacing slot liners with surface coating (essentially paint)
%thermal interface to slot walls; higher temperatures (materials, PEEK, ...)

\subsection{Speed and AC Effects}
The automotive industry faces the same high-frequency ac effects that challenged and almost broke bar windings in their early days. Furthermore, high turn ratios and large flux linkage levels are trickier than with thin wires. The early bar-wound traction motors therefore still used rather low speeds, which allow fitting more series-connected poles into the machine for sufficient inductance while concurrently controlling the electrical frequency. Instead, these motors rather exploited the torque density potential associated with the high slot-fill. 

Re-learning earlier tricks and the development of new technology, particularly more confidence and experience with winding schemes, however, has led to a constant increase of the possible rotational speed of bar-wound devices for series-manufactured vehicles (Fig.\ \ref{fig:trend}). This trend is known from wire-wound automotive drives and is with only small delay matched by hairpin machines starting with the earliest machines in the Chevrolet Voltec with only few thousand rotations per minute to latest sports cars. The higher speeds in combination with the increase in voltage, most recently with the success of 800~V drive trains, has enabled a notable increase in power density (Tables \ref{table:modern} and \ref{table:modern1}).

%Like other automotive machines, also bar-wound machines are pushed towards higher speeds for the same size, with currently approaching 30,000~r/min, which increases the overall power density. This trend towards higher speeds can be clearly observed for more than ten years and started with the earliest use of bar-wound machines in the hybrid-electric Chevrolet vehicles with only few thousand rotations per minute to latest battery-electric sports cars (Fig.\ \ref{fig:trend}).

As in wire-wound machines, the mechanical stability of the rotor (particularly for permanent magnet machines) and unequal thermal expansion (for induction machines) challenge this increase in speed. However, in contrast to wire-wound machines, %there is typically an even more severe limitation: the increasing frequency and the associated ac loss in the bars.
the major limitation is the increasing electrical frequency and the associated ac loss in the bars.
A further reduction of the number of poles leaves only limited potential as many current designs for electrical vehicles have anyway only four poles (two pole pairs) left. Accordingly, the increasing speed is closely related to efforts mitigating the ac effects.

A large body of ongoing research for managing the ac effects aims at winding schemes and the bar shapes. Increasing the number of bars per slot and short-pitching in combination with alternating the bars of different phases in a slot has been known to reduce ac effects for a long time as indicated above. However, winding schemes and smart end-turn patterns still offer a considerable potential. The early winding schemes with more bars per slot used a fundamental pattern and repeated it for the remaining space. If the copies are shifted against the mother pattern, the outcome results in a short-pitched configuration, although bars from the same strand and thus current phase might aggregate so as to lie side by side where they influence each other through proximity. Smarter pin transposition groups that mix bars from various phases and fully interwoven patterns for six, eight, and more bars per slot with as few different parts as possible---to limit the number of tools if no computer numerical control (CNC) benders are used (Fig.\ \ref{fig:manufacturingCNC})---might get established soon.

As has been known for a long time, the space close to the air gap is most sensitive to loss. Transversal leakage flux passing through a slot driven by stator flux that does not reach the airgap tends to increase closer to the air gap, particularly if the teeth approach magnetic saturation, and can induce eddy currents in bars with large radial size. Likewise rotor leakage flux can enter slots from the airgap, as hinted in Section IV, and pass through the first bars radially, where they increase loss, particularly if they span the entire slot width. To avoid the general trade-off between slot fill and ac loss, windings schemes with bars of varying cross sections are under research \cite{9254278,9270927,9541356,9509846,9152417}. Furthermore, the shape of the bars may be adjusted to the flux lines, which can be achieved with cold rolling.
Magnetic wedges or even magnetic coating to close the slot but with either generally low permeability or early saturation to not shorten the stator flux can further absorb rotor stray flux and prevent it from reaching into slots \cite{DE102019109728}.

%But for high voltage applications, rectangular wire is new. GM first introduced the rectangular wire innovation in their 2 Mode hybrids in 2008, and has used the technology ever since. Most recently, the Chevy Bolt features the technology in its highest art, as shown in section photo. Also, Toyota features rectangular wire construction in the Prius C (Aqua) hybrid and their latest generation Prius. Ferrari has used it, and Honda is said to introduce it into their latest electrified vehicles. Other OEMs are soon following.

Aside from OEMs, one can see rectangular wire know-how spread into the EV motor supply chain.  Several tier 1 motor manufacturers are highly skilled in the art of bar windings already, such as Borg Warner (formerly Remy), Hitachi, LG, Denso, or Magneti Marelli, while others are quickly catching up. Tier 2 supplier companies manufacturing high-quality rectangular profile wire, insulation systems, and manufacturing equipment can be found in North America, Europe, and throughout Asia. 

The benefits of bar-wound motor designs are tangible, whereas solutions to mitigate the drawbacks are developing rapidly. While electric vehicles are coming to the fore in the top and mid-segment passenger vehicle market, commercial vehicles such as trucks are on the horizon. Bar windings offer both high performance and economic manufacturing. Wire-wound motors might well be relegated to lower performance and special applications. %Bar-wound motor technology has come into its own as the preferred mainstream motor technology for series electric vehicles.

\section{Conclusion}

\begin{comment}
\begin{itemize}
    \item Hairpin and form windings developed earlier
    \item maybe too early
    \item 
\end{itemize}
\end{comment}

Bar windings, particularly hairpin, have rapidly gained dominance in traction drives of recent vehicles. On the one hand, significant advancements have lead to highly-efficient manufacturing processes which not only match the automotive production standards: fully automated, resource efficient with short lead times \cite{7323198,Weigelt2018,8658353}. On the other hand, bar windings offer additional benefits that are essential in this challenging industry \cite{Acatech2010}, most importantly larger slot fill factor (up to $\approx{}80\%$) as well as more compact winding overhang enabling improved power density and/or efficiency, better heat dissipation, smaller package space, as well as longer insulation life  \cite{7008430,7604085,8658353,8766938}.

Although this technology appears to be brand-new and to take the field of car propulsion motors by storm, a closer look at older literature and historic engineering artifacts discloses that bar windings including hairpins were developed way earlier and in greater depths than most engineers even from the field may have expected. Potentially, this technology was even developed too early: Many problems and effects were not understood sufficiently at a time when motors themselves were still a relatively young technology with many open questions. This article discussed the ac effects as an example. Furthermore, a number of features of bar windings are hailed nowadays, particularly fully automated assembly and better thermal conditions as well as superior cooling approaches possible with bar windings, which were not needed back then.

Nevertheless, the consequence is that many discoveries and solutions were forgotten, while the technology was not required in many applications and only partly survived in a niche.
It turns out, however, that the rapid development of this seemingly new technology after a long disappearance has led to numerous re-inventions and even patent applications that were initially invented already decades or even more than a century ago. Bar windings, including segmented windings out of hairpins, might be an exemplary field where a closer look at the historic developments turns uncovers that many present challenges already have solutions or inspirational approaches.

% \section*{Acknowledgment}

% The authors would like to thank...

% Can use something like this to put references on a page
% by themselves when using endfloat and the captionsoff option.
\ifCLASSOPTIONcaptionsoff
  \newpage
\fi

% trigger a \newpage just before the given reference
% number - used to balance the columns on the last page
% adjust value as needed - may need to be readjusted if
% the document is modified later
%\IEEEtriggeratref{8}
% The "triggered" command can be changed if desired:
%\IEEEtriggercmd{\enlargethispage{-5in}}

% references section

% can use a bibliography generated by BibTeX as a .bbl file
% BibTeX documentation can be easily obtained at:
% http://mirror.ctan.org/biblio/bibtex/contrib/doc/
% The IEEEtran BibTeX style support page is at:
% http://www.michaelshell.org/tex/ieeetran/bibtex/
%\bibliographystyle{IEEEtran}
% argument is your BibTeX string definitions and bibliography database(s)
%\bibliography{IEEEabrv,../bib/paper}
%
% <OR> manually copy in the resultant .bbl file
% set second argument of \begin to the number of references
% (used to reserve space for the reference number labels box)
%\bibliographystyle{IEEEtranTIE}
%\bibliography{IEEEabrv,references}

% \section{Conclusion}
% The conclusion goes here.

%%%%%%%%%%%%%%%%%%%%%%%%%%%%%%%%

% Generated by IEEEtran.bst, version: 1.12 (2007/01/11)

% that's all folks
\end{document}